\documentclass{jfm}
\usepackage{soul}
\usepackage{amsmath,amssymb,capt-of,ifthen,calc,graphicx,bm} \usepackage{natbib}

\let\realverbatim=\verbatim \let\realendverbatim=\endverbatim
\renewcommand\verbatim{\par\addvspace{6pt plus 2pt minus 1pt}\realverbatim}
\renewcommand\endverbatim{\realendverbatim\addvspace{6pt plus 2pt minus 1pt}}
\makeatletter \newcommand\verbsize{\@setfontsize\verbsize{10}\@xiipt}
\renewcommand\verbatim@font{\verbsize\normalfont\ttfamily} \makeatother

\ifCUPmtlplainloaded \else \checkfont{eurm10} \iffontfound
\IfFileExists{upmath.sty} {\typeout{^^JFound AMS Euler Roman fonts on the
    system, using the 'upmath' package.^^J}%
  \usepackage{upmath}} {\typeout{^^JFound AMS Euler Roman fonts on the system,
    but you don't seem to have the}%
  \typeout{'upmath' package installed. jfm.cls can take advantage of these
    fonts,^^Jif you use 'upmath' package.^^J}%
  \providecommand\mathup[1]{##1}%
  \providecommand\mathbup[1]{##1}%
} \else \providecommand\mathup[1]{#1}%
\providecommand\mathbup[1]{#1}%
\fi \fi

\ifCUPmtlplainloaded \else \checkfont{msam10} \iffontfound
\IfFileExists{amssymb.sty} {\typeout{^^JFound AMS Symbol fonts on the system,
    using the 'amssymb' package.^^J}%
  \usepackage{amssymb}%
     \let\geq=\geqslant }{}
\fi \fi

\ifCUPmtlplainloaded \else \IfFileExists{amsbsy.sty} {\typeout{^^JFound the
    'amsbsy' package on the system, using it.^^J}%
  \usepackage{amsbsy}} {\providecommand\boldsymbol[1]{\mbox{\boldmath $##1$}}}
\fi

\newsavebox{\astrutbox} \sbox{\astrutbox}{\rule[-5pt]{0pt}{20pt}}

\usepackage{fancyhdr}
\usepackage{hyperref}

\title[Journal of Fluid Mechanics] {Accuracy analysis of high-order lattice
  Boltzmann models for rarefied gas flows\footnote{Corresponding author: yonghao.zhang@strath.ac.uk }}

\author{Jianping Meng and Yonghao Zhang}

\affiliation{Department of Mechanical Engineering, University of Strathclyde, Glasgow G1
  1XJ, UK}

\begin{document}

\label{firstpage}
\maketitle
\thispagestyle{empty}

\begin{abstract}

  In this work, we have theoretically analyzed and numerically evaluated the accuracy of high-order lattice Boltzmann (LB) models for capturing non-equilibrium effects in rarefied gas flows. In the
  incompressible  limit, the LB equation is proved to be equivalent to the linearized Bhatnagar-Gross-Krook (BGK)
  equation. Therefore, when the same Gauss-Hermite quadrature is used, LB method closely assembles the discrete velocity method (DVM). In addition, the order of Hermite expansion for the equilibrium distribution function is found not to be correlated with
  the approximation order in terms of the Knudsen number to the BGK equation, which was
  previously suggested by \cite{2006JFM...550..413S}. Furthermore, we have
  numerically evaluated the LB models for a standing-shear-wave
  problem, which is designed specifically for assessing model accuracy by
  excluding the influence of gas molecule/surface interactions at wall
  boundaries. The numerical simulation results confirm that the high-order terms in the discrete equilibrium distribution
function play a negligible role. Meanwhile, appropriate Gauss-Hermite quadrature has the most significant effect on whether LB models can describe the essential flow physics of rarefied gas accurately. For the same order of the Gauss-Hermite
  quadrature, the exact abscissae will also modestly influence numerical accuracy. Using
  the same Gauss-Hermite quadrature, the
  numerical results of both LB and DVM methods are in excellent agreement for flows across a broad range of the Knudsen numbers, which confirms that the LB simulation is similar to the DVM process. Therefore, LB method can offer flexible models suitable for simulating continuum flows at Navier Stokes level and rarefied gas flows at the linearized Boltzmann equation level.
 
\end{abstract}

\section{Introduction}
Rarefied gas flows have recently attracted significant research interest due to
the rapid development of micro/nano-fluidic technologies. Gaseous transport in
micro/nano devices is often found to be non-equilibrium, and non-equilibrium
phenomena have not yet been well understood
\cite[][]{annurev.fluid.30.1.579}. The conventional theory to describe gas flows
is the Navier Stokes equations, which assume that the fluid is in a quasi-equilibrium state. However, for non-equilibrium flows, the Navier Stokes equations break down because that the molecular nature of the gas strongly
affects the bulk flow behavior i.e. the gas can no longer be regarded as a
fluid continuum. Whether gas flows are in local equilibrium or not can be
classified by the non-dimensional Knudsen number, Kn, defined as the ratio of
mean free path and the device characteristic length scale. The Navier Stokes
equations with no-velocity-slip wall boundary condition are only appropriate
when $Kn<0.001$. However, gas flows in micro/nano-fluidic devices are often in
the slip flow regime ($0.001<Kn<0.1$) or the transition flow regime ($0.1<Kn<10$). In
these regimes, the gas flow cannot properly be described as a continuous
flow, nor as a free molecular flow. In practice, most devices
operate with a range of Knudsen numbers in different parts of the device; this
makes it even more difficult to develop a generalized flow model.

Direct simulation Monte Carlo (DSMC) methods and direct numerical simulation of
the Boltzmann equation can provide accurate solutions for rarefied gas
flows. However, these are computationally intractable for 3D flow systems, and
impractical with the current computer technology, especially for the low speed gas
flows usually encountered in micro/nano-systems. Statistically, one needs to take significantly more 
samples of the flow field at any point for the DSMC method to resolve flows with low Mach numbers. The
direct simulations based on the Boltzmann equation requires significant
computational resources for integrating the velocity space ranging from
$-\infty$ to $+\infty$. In addition, it is usually difficult to solve the full
Boltzmann equation directly via either numerical or analytical methods.
 
Meanwhile, the continuum methods beyond the Navier Stokes level have failed to
produce satisfactory results for gas flows in the transition flow regime
\cite[][]{Lockerby2008}.  It is well-known that continuum expressions for the viscous stress and
heat flux in gases may be derived from the fundamental Boltzmann equation via
either a Kn-series solution (known as the Chapman-Enskog approach) or by an
expansion of the distribution function as a series of Hermite tensor polynomials
\cite[][]{1991mtnu.book.....c}. To first order (i.e. for near-equilibrium flows) both approaches yield the
Navier Stokes equations. However, the solution methods can be continued to
second and higher orders, incorporating more and more of the salient
characteristics of a non-equilibrium flow. The classical second-order stress and
heat flux expressions are the Burnett equations (from the Chapman-Enskog
approach), and the Grad 13-moment equations (from the Hermite polynomial method)
\cite[][]{1991mtnu.book.....c}. These can be seen as corrections to the Navier Stokes constitutive relations to make them more appropriate to continuum-transition flows. However, different physical interpretations of the solution methods at second and higher
orders have recently led to a variety of sets of
equations, including the Bhatnagar-Gross-Krook (BGK)-Burnett\cite[][]{2004jfm...503..201b},
Eu \cite[][]{physreve.70.016301}, augmented Burnett \cite[][]{1993aiaaj..31.1036z}, and
regularized moment (R13)\cite[][]{struchtrup:2668} equations. While each purports to be the proper
high-order correction to the stress and heat flux (there is no disagreement
about the form of the Navier Stokes equations at first-order), none of these models are
satisfactory \cite[][]{Lockerby2008}. In addition, these models suffer unknown
additional boundary conditions at solid walls to appropriately reflect gas molecule/wall surface interactions.
 
The lattice Boltzmann (LB) approach offers an alternative method for rarefied
gas flow simulations. Historically, the LB model was evolved from the
lattice-gas automata (LGA) for mimicking the Navier Stokes hydrodynamics \cite[see][and references therein]{qian1992,1998anrfm..30..329c,1992phr...222..145b}.
Over the past two decades, the LB method has been developed to provide accurate
and efficient solutions for continuum flow simulations as the validity of the model
can be ensured by the Chapman-Enskog expansion. Due to its kinetic nature, the
LB model has distinct advantages over the continuum computational methods,
including easy implementation of multi-physical mechanisms and the boundary
conditions for fluid/wall interactions. The potential of LB models for simulating rarefied gas flows have been demonstrated \cite[e.g.][]{zhang:047702,Toschi2005,sbragaglia:093602,Sbragaglia2006,tang:046701,zhang:046704,2006JFM...550..413S,
Ansumali2007, Kim20088655, yudistiawan:016705}. 

Recently, the LB models were shown to be able to be derived
systematically from the Boltzmann-BGK equation based on the Hermite expansion
 \cite[see][]{PhysRevLett.80.65, PhysRevE.56.6811, 2006JFM...550..413S}. This
creates another theoretical foundation different from the
Chapman-Enskog expansion, so that higher-order LB approximations to
the Boltzmann-BGK equation beyond the Navier Stokes level can be constructed
by using the high-order Hermite expansion with appropriate quadratures. This indicates that
high-order LB models have the potential to capture
non-equilibrium effects in rarefied flows. In addition to the systemic framework of
constructing LB models, \cite{2006JFM...550..413S} also established the link between the orders of 
Hermite polynomials and Chapman-Enskog expansion. The authors concluded that the order of Hermite expansion is responsible for obtaining correct velocity moments. The precise relation among the orders of Hermite
expansion, Chapman-Enskog expansion and velocity moments was described by Eq.(4.7) in \cite{2006JFM...550..413S}. For instance, the third-order expansion is required for accurate pressure tensor and momentum at the Navier Stokes level. These conclusions are key to constructing appropriate LB models for non-equilibrium gas flows. However, the numerical simulations do not support these conclusions. In contrast, the simulation data showed that the higher order terms in the equilibrium distribution function have negligible influence for low speed rarefied flows \cite[] {Kim20088655}. This indicates that the Hermite expansion order is not related to the order of Chapman-Enksog expansion, in contrary to the theoretical conclusions drawn by \cite{2006JFM...550..413S}.

In this work, we aim to answer this question whether the Hermite expansion order is important for the LB method, as it is for the Grad's moment method, to capture non-equilibrium effects in rarefied flows, especially at micro/nano-scales. Furthermore, we will analyze theoretically and numerically the mechanisms that are important in constructing high-order LB models for rarefied gas dynamics. We will discuss the differences between the approaches of \cite{2006JFM...550..413S} and  Grad's moment method. To help us to understand the modeling
capability of the LB method for rarefied gas dynamics, we will also analyze the
similarities and differences between the LB method and the discrete velocity
method (DVM) of solving the BGK equation. In particular, we will prove that the Hermite expansion order is not important for the flows that the linearized BGK equation can accurately describe. Since the important nonlinear constitutive relations in the Knudsen layer are still not captured satisfactorily \cite[][]{tang:046701}, our numerical analysis will be based on a standing-shear-wave problem specifically designed by \cite{Lockerby2008} to exclude the effect of gas molecule/wall interactions, so we can concentrate on the model capabilities.



\section{Lattice Boltzmann simulation of rarefied gas flows}

\subsection{Lattice Boltzmann equation}

Although the LB models were originally developed from LGA, the link to the
kinetic theory has late been established by
\cite{PhysRevE.55.R6333,PhysRevE.56.6811,PhysRevLett.80.65,2006JFM...550..413S}.
Consequently, the LB approach may be considered as a special finite difference
scheme of solving the Boltzmann-BGK equation \cite[][]{Luo200063}.  This
theoretical link indicates that the LB methodology may provide a reasonable approximation to the Boltzmann-BGK equation. The central question
is how accurate the LB models can capture non-equilibrium effects in rarefied
gas dynamics. To answer this question, we will revisit the derivation process
of LB models from the Boltzmann-BGK equation proposed by \cite{2006JFM...550..413S} and we will emphasize on the model capability in describing rarefied gas flows.

The original Boltzmann-BGK equation can be written as:
\begin{equation}
  \label{dbgk}
  \frac{\partial f}{\partial t} + \bm \xi \cdot \nabla f + \bm g \cdot \nabla_\xi f=
  -\frac{p}{\mu}\left(f-f^{eq} \right),
\end{equation}
where $f$ denotes the distribution function, $\bm \xi$ the phase velocity, $p$
the pressure, $\bm g$ the body force and $\mu$ the gas viscosity. Using the
  well-known Chapman-Enskog expansion, the collision frequency can be
  represented by the ratio of pressure and gas viscosity, which is convenient to obtain the Knudsen number
  definition consistent with that of hydrodynamic models. Without losing
generality, we define the following non-dimensional variables:
\begin{eqnarray}
  & \displaystyle \hat{\bm r}=\theta \bm r, \hat{\bm u}=\frac{\bm u}{\sqrt{R T_0}},\hat{t}=\theta \sqrt{R T_0} t,
  \nonumber\\
  & \displaystyle \hat{\bm g}=\frac{\bm g}{\theta R T_0}, \hat{\phi}=\frac{\phi}{\theta \sqrt{R T_0}},
  \hat{\bm \xi}= \frac{\bm \xi}{\sqrt{R T_0}}, \hat{T}=\frac{T}{RT_0}, &
  \label{nondime}
\end{eqnarray} 
where $\bm u$ is the macroscopic velocity, $R$ the gas constant, $T$ the gas
temperature, $T_0$ the reference temperature, $\bm r$ the spatial position and $\theta$ the inverse of the characteristic
length of the flow system. The symbol {\it hat}, which denotes a dimensionless
value, will hereinafter be omitted. We define the Knudsen number using macroscopic properties as below:
\begin{equation}
  \label{KN}
  Kn=\frac{\theta \mu \sqrt{RT_0}}{p}.
\end{equation}
By using these non-dimensional variables, the non-dimensional form of the
Boltzmann-BGK equation becomes
\begin{equation}
  \label{bgk}
  \frac{\partial f}{\partial t} + \bm \xi \cdot \nabla f + \bm g \cdot \nabla_\xi f=
  -\frac{1}{Kn}\left(f-f^{eq} \right),
\end{equation}
where the Maxwell distribution in $D$-dimensional Cartesian coordinates can be written as
\begin{equation}
  f^{eq}=\frac{\rho}{(2\pi T)^{D/2}} \exp\left[\frac{-(\bm \xi-\bm u)^2}{2 T}\right].
\end{equation}
From the non-dimensional format of Eq.(\ref{bgk}), we can clearly see the relationship between the relaxation time and the mean free path (i.e.
Knudsen number), which plays a key role in LB simulation of rarefied gas flows \cite[e.g.][]{zhang:047702}.    

To discretize the velocity space, we project the distribution function onto a functional space spanned by the
orthogonal Hermite basis:
\begin{equation}
\label{approxf}
  f(\bm r,\bm \xi,t) \approx f^{N}(\bm r,\bm \xi,t)=\omega(\bm \xi) \sum^N_{n=0}\frac{1}{n!}\bm a^{(n)}(\bm r,t) \bm \chi^{(n)}(\bm \xi),
\end{equation}
where $\chi^{(n)}$ is the $n$th order Hermite polynomial. The weight function
$\omega(\xi)$ is given by
\begin{equation}
  \omega(\bm \xi)=\frac{1}{(2 \pi)^{D/2}} \text{e}^{-\xi^2/2},
\end{equation}
and the coefficients $\bm a^{(n)}$ are
\begin{equation}
  \label{an}
  \bm a^{(n)}=\int f \bm \chi^{(n)} d\bm \xi \approx \int f^{(N)}  \bm \chi^{(n)} d\bm \xi =\sum^d_{\alpha=1} \frac{w_\alpha}{\omega(\bm \xi_\alpha)}f^{(N)}(\bm r, \bm
  \xi_\alpha,t)\bm \chi^{(n)}(\bm \xi_\alpha).
\end{equation}
 The coefficient $a_{eq}^{(n)}$ for the equilibrium distribution is
\begin{equation}
  \bm a_{eq}^{(n)}=\int f^{eq}\bm \chi^{(n)} d\bm \xi. 
\end{equation}
where $w_\alpha$ and $\bm \xi_\alpha$, $a=1,\cdots,d$, are the weights and
abscissae of a Gauss-Hermite quadrature of degree $\geq 2N$
respectively. Herein, the distribution function is approximated by the first $N$
Hermite polynomial. Using the derivation relation, the body force term  $F(\bm r,\bm
\xi,t) = \bm g \cdot \nabla_\xi f$   can be approximated as 

\begin{equation}
  F(\bm r,\bm \xi,t) =w \sum^N_{n=1}\frac{1}{(n-1)!}\bm g \bm a^{(n-1)}\bm \chi^{(n)}(\bm \xi_\alpha).
\end{equation}

As an example, the second order approximation of the equilibrium distribution and the body force are:
\begin{equation}
\label{demoeq}
f^{eq} \approx \omega(\bm \xi) \rho \left\{1+\bm \xi \cdot \bm u +
    \frac{1}{2}\left[(\bm \xi \cdot \bm u)^2-u^2+(T-1)(\xi^2-D)\right] \right\},
\end{equation}

\begin{equation}
F(\bm r,\bm \xi,t)  \approx \omega(\bm \xi) \rho \left\{\bm g \cdot \bm \xi
  +(\bm g \cdot \bm \xi )(\bm u \cdot \bm \xi)-\bm g \cdot \bm u \right\},
\end{equation}
where $T$ should be set to unity for isothermal problems and $\rho$ is constant for incompressible problems.

An appropriate Gauss-Hermite quadrature,
see the Appendix in \cite{2006JFM...550..413S} for a list of quadratures, can be
chosen to evaluate the integral to obtain $a^{(n)}$. Consequently, Eq.(\ref{bgk})
can be discretized as 
\begin{equation}
  \label{lbgk}
  \frac{\partial f_\alpha}{\partial t}+\bm \xi_\alpha \cdot \nabla f_\alpha =-\frac{1}{Kn}\left(f_\alpha-f_\alpha^{eq}\right)+g_\alpha,
\end{equation}
where $f_\alpha=w_\alpha f(\bm r,\bm \xi_\alpha,t)/\omega(\bm \xi_\alpha)$, $f_\alpha^{eq}=w_\alpha f^{eq}(\bm
r,\bm \xi_\alpha,t)/\omega(\bm \xi_\alpha)$ and $g_\alpha=w_\alpha \bm F(\bm r,\bm \xi_\alpha,t)/\omega(\bm \xi_\alpha)$.  We have obtained the lattice Boltzmann equation, i.e. Eq.(\ref{lbgk}), by discretizing Eq.(\ref{bgk}) in the velocity space.

\subsection{Numerical schemes, Knudsen number and relaxation time}
An appropriate numerical scheme is now required to solve Eq.(\ref{lbgk}). If a
finite difference scheme is chosen, one can obtain the so-called finite
difference lattice Boltzmann model. In particular, when the first-order upwind
finite-difference scheme is chosen, one can obtain the standard form of LB
model:
\begin{equation}
  \label{slb}
  f_\alpha(\bm r+\bm \xi_\alpha \delta t,t+\delta t)-f(\bm r,t)= -\frac{\delta t}{Kn}(f_\alpha-f_\alpha^{eq})+\delta t g_\alpha,
\end{equation}
where the relationship between the relaxation time $\tau$ and the Knudsen number
is established naturally i.e. $\tau=kn/\delta t$. For continuum
flows where the Navier Stokes equations are valid, the above first-order scheme
can become effectively second-order accurate in both space and time by simply
replacing the non-dimensional relaxation time $\tau$ with $\tau-0.5$ \cite[see][]{Reider1995459,Sterling1996}. In doing so, the second order discretization
error can be absorbed into an artificial viscosity. Therefore, this simple but
accurate scheme has been widely used to simulate flows at the Navier Stokes
level. Since any LB model intended to simulate rarefied gas dynamics beyond the
Navier Stokes level needs to recover the Navier Stokes equation at small Knudsen
number, i.e. $Kn\rightarrow 0$, this first-order scheme with correction has been commonly used in LB
simulation of rarefied gas flows \cite[e.g.][]{Nie2002,tang:046701}
for rarefied gas problems. However, the artificial viscosity has only corrected
the momentum transfer to the second order, which is only appropriate for the
Navier-Stokes hydrodynamics. This correction will lead to inconsistency for the
transfer of the other higher-order moments, which are essential for capturing
non-equilibrium effects in rarefied gas flows. Therefore, the dilemma is that we
need correction on the relaxation time to recover the Navier Stokes
hydrodynamics appropriately when the Knudsen number is close to zero where the
high-order moments are not important. Meanwhile, we should not have this
correction for the higher-order moments which are more important to rarefied
flows. In deed, the simulation will diverge when the Knudsen number is
approaching to 0.5 if no relaxation time correction is introduced. The reason is
that it goes beyond the stability regime of the relaxation scheme.
\cite{lim:2299} suggested to use the correction when the Knudsen number is less
than 0.5 and switch to no correction when the Knudsen number is larger
than 0.5. However, it will lead to inconsistency at the Knudsen number around
0.5 which is the most important flow regime in micro/nano-fluidic devices. The
above first-order upwind scheme should not be used for simulating the gas flows with
finite Knudsen numbers.

To resolve this problem inherited from the standard LB method, we should not use
the artificial viscosity to achieve correct physics at the Navier Stokes
level. We propose to discretize Eq.(\ref{lbgk}) using a numerical scheme with second-order accuracy,
which was first used by \cite{He1998282} for thermal flow simulation at the Navier Stokes level:
\begin{eqnarray}
  f_\alpha(\bm r+\bm \xi_\alpha \delta t,t+\delta t)-f(\bm r,t) & = & -\frac{\delta t}{2kn}\left[f_\alpha(\bm r+ \bm \xi_\alpha \delta t,t+\delta t) -f_\alpha^{eq}(\bm r+ \bm \xi_\alpha
    \delta t,t+\delta t)\right] \nonumber \\
  &  & -\frac{\delta t}{2kn}\left[f_\alpha(\bm r,t)-f_\alpha^{eq}(\bm r,t)\right] \nonumber \\ 
  & & +\frac{\delta t}{2} \left[g_\alpha(\bm r+\bm \xi_\alpha \delta t,t+\delta t)+ g_\alpha(\bm r,t)\right].  
\end{eqnarray}
By introducing
\begin{equation}
  \tilde{f}_\alpha=f_\alpha + \frac{\delta_t}{2 kn} (f_\alpha-f_\alpha^{eq})-\frac{\delta_t}{2} g_\alpha,
\end{equation}
the above implicit scheme can be written as
\begin{equation}
\label{2order}
  \tilde{f}_\alpha( \bm r+\bm \xi_\alpha \delta_t,t+\delta_t)-\tilde{f}_\alpha(\bm r,t)=
  -\frac{\delta_t}{Kn+0.5\delta_t}\left[\tilde{f}_\alpha(\bm r,t)-f_\alpha^{eq}(\bm r,t) \right]
  +\frac{Kn  g_\alpha \delta_t}{Kn+0.5\delta_t},
\end{equation}
with
\begin{equation}
  \rho = \sum_\alpha \tilde{f}_\alpha,
\end{equation}

\begin{equation}
  \rho \bm u =\sum_\alpha \bm \xi_\alpha \tilde{f}_\alpha+\frac{\rho  \bm g \delta_t}{2}.
\end{equation}
Therefore, the viscosity is now $\tau RT$ rather than $(\tau-0.5)RT$. Most
importantly, the same relation between the relaxation time and the mean free
path can be used for the transfer of any order moments.

\subsection{High-order lattice Boltzmann models}
Although the construction of LB models based on the Hermite polynomials is
straightforward, the Hermite polynomials higher than the third order give
irrational roots. The integer stream velocity is an essential feature of LB
models, i.e. the simple and efficient ``stream-collision'' mechanism. So
high-order LB models, which have non-integer discrete velocities, will need
additional effort, such as point-wise interpolation \cite[][]{He1996}. Therefore, they
essentially become off-lattice discrete velocity method for solving the kinetic
Boltzmann equation, which will increase the computational cost dramatically and
introduce extra numerical error. \cite{2006JFM...550..413S} suggested a method
for searching abscissae on the grid points of Cartesian coordinates to construct
high-order LB models with integer discrete velocities. The examples are D2Q17 and
D2Q21 models given by \cite{2006JFM...550..413S} and \cite{Kim20088655} (note,
we follow the conventional terminology for the LB models as first introduced by
\cite{qian1992} dubbed as DnQm model i.e. n dimensional model with m discrete
velocities). Furthermore, \cite{Chikatamarla2006} proposed an alternative method
to seek rational-number approximation to the rations of the Hermite roots based
on the relation between the entropy and the roots of Hermite
polynomials. They also proposed the higher-order LB models with integer discrete
velocity, such as D2Q16 and D2Q25 models. The above high-order LB models with
integer stream velocities will be numerically examined in this work and the
details are listed in Table.(\ref{model}).

\begin{table}

  \begin{center}
    \begin{tabular}{lllll}
      \hline
      \hline
      Quadrature & k  &   $\xi_{\alpha}$  & \ $w_{\alpha} $ & \\
      \hline
      D2Q9   & 1 & (0,0) & 4/9  &  \\
      & 4 & $(\sqrt{3},0)_{FS}$ & 1/9 & \\
      & 4 & $(\pm \sqrt{3},\pm \sqrt{3})$ & 1/36 & \\
      \hline
      D2Q16 &  4 & $(\pm m,\pm m)$ & $W_{\pm m}^2 $ & $ m=1,n=4$\\
      & 4 & $(\pm n,\pm n)$ & $W_{\pm n}^2$ & $W_{\pm m}=\frac{m^2-5n^2+\sqrt{m^4-10n^2m^2+n^4}}{12(m^2-n^2)}$\\
      & 4 & $(\pm m,\pm n)$ & $W_{\pm m}W_{\pm n}$ & $W_{\pm n}=\frac{5m^2-n^2-\sqrt{m^4-10n^2m^2+n^4}}{12(m^2-n^2)}$ \\
      & 4 & $(\pm n,\pm m)$ & $W_{\pm m}W_{\pm n}$ & $T_0=(m^2+n^2+\sqrt{m^4-10n^2m^2+n^4})/6$\\
      \hline
      D2Q17 & 1 & (0,0) & $(575+193\sqrt{193})/8100$ &  \\
      & 4 & $(r,0)_{FS}$ & $(3355-91\sqrt{193})/18000$& $r^2=(125+5\sqrt{193})/72$\\
      &  4 & $(\pm r,\pm r)$ & $(655+17\sqrt{193})/27000$ & \\
      &  4 & $(\pm 2r,\pm 2r)$& $(685-49\sqrt{193})/54000 $ & \\
      & 4 & $(3r,0)_{FS} $ & $(1445-101\sqrt{193})/162000 $ &\\
      \hline

      D2Q21 & 1 & $(0,0)$ & $ 91/324 $ &  \\
      & 4 & $(r,0)_{FS}$ & $ 1/12 $& $r^2=3/2$\\
      &  4 & $(\pm r,\pm r)$ & $ 2/27 $ & \\
      &  4 & $(\pm 2r,0)_{FS} $& $ 7/360 $& \\
      &  4 & $(\pm 2r,\pm 2r)$& $ 1/432 $ & \\
      & 4 & $(3r,0)_{FS} $ & $1/1620 $ &\\
      \hline
      D2Q25 & 1 & $(0,0)$ & $W_0^2$ &  $m=3$, $n=7$\\
      & 4 & $(m,0)_{FS}$ & $W_{\pm m}W_0$ & $W_0=\frac{-3m^4-3n^4+54m^2n^2-(m^2+n^2)D_5}{75m^2n^2}$\\
      & 4 & $(n,0)_{FS}$ & $W_{\pm n}W_0$ & $W_{\pm m}=\frac{9m^4-6n^4-27n^2m^2+(3m^2-2n^2)D_5}{300m^2(m^2-n^2)}$\\
      & 4 & $(\pm m,\pm m)$ & $W_{\pm m}^2$ & $W_{\pm n}=\frac{9n^4-6m^4-27n^2m^2+(3n^2-2m^2)D_5}{300n^2(n^2-m^2)}$\\
      & 4 & $(\pm n,\pm n)$ & $W_{\pm n}^2$ & $T_0=(3m^2+3n^2+D_5)/30$\\
      & 4 & $(\pm m,\pm n)$ & $W_{\pm m}W_{\pm n}$ & $D_5=\sqrt{9m^4-42n^2m^2+9n^4}$\\
      & 4 & $(\pm n,\pm m)$ & $W_{\pm m}W_{\pm n}$ & \\
      \hline
      \hline
    \end{tabular}
  \end{center}
  \caption{The quadratures of five LB models where k is the number of discrete velocities with the same velocity magnitude, the subscript $FS$ denotes a fully symmetric set of points, and $w_\alpha$ are the weights. The quadrature accuracy is fifth-order for the D2Q9 model, seventh-order for the D2Q16, D2Q17 and D2Q21 models, and ninth-order for the D2Q25 model. The details of D2Q17 and D2Q21 models can be found in \cite{2006JFM...550..413S,Kim20088655} while the D2Q16 and D2Q25 models are discussed in \cite{Chikatamarla2006}.
  }
  \label{model}
\end{table}

Based on the above model construction procedure, the accuracy of LB models
depends on three level of approximations. Firstly, it depends on the accuracy of
the numerical scheme for solving Eq.(\ref{lbgk}). As we have demonstrated, the commonly used first-order upwind scheme will lead to incorrect physics for rarefied flow\textbf{s}. Our second-order numerical scheme given by Eq. (\ref{2order}) is essential to capture
non-equilibrium effects accurately. Secondly, the order of the Hermite
expansion was considered to be important to obtain the correct moments
\cite[][]{2006JFM...550..413S}. Thirdly, the Gauss-Hermite quadrature accuracy
should be sufficiently high so that the integration of Eq.(\ref{an}) can be
evaluated accurately. Therefore, the term {\it higher-order} LB models here
refer to the LB models with high-order of Hermite expansion and Gauss-Hermite quadrature in comparison with the standard LB model.


\section{Lattice Boltzmann, moment and discrete velocity methods}

\subsection{Comparison of Grad's moment method and lattice Boltzmann method}

Similar to the Grad's method for deriving higher order continuum systems (e.g,
  Grad 13-moment equations), using the Hermite expansion to
  approximate the Boltzmann-BGK equation can lead to the LB equation, i.e. Eq.(\ref{lbgk}). However, the major difference is that
  LB models are always staying at the kinetic level, i.e. solving the kinetic equation - Eq.(\ref{lbgk}),  while the Grad's method will produce a set of continuum equations. The basic idea of Grad's method is to use the truncated Hermite polynomials to approximate the full Boltzmann (or Boltzmann-BGK) equation. Due to the unique feature of Hermite polynomial, the  moments of up to the chosen truncation order can be described accurately
  by the derived macroscopic moments systems. In contrary, the only explicit
  effect of the truncation on the LB models is on the approximation of the equilibrium distribution
  function and the body force, while the Grad's moment equations do not approximate the equilibrium distribution function. 
  
 Although the order of Hermite expansion determines the accuracy level of the moment model, which is not the same for the LB models.  Essentially, the LB equation i.e. Eq.(\ref{lbgk}) is similar to any model equation which is to simplify the full Boltzmann equation. The kinetic process, i.e. gas molecules relaxing to the equilibrium state through collisions, is still the same. Therefore, the LB method is very close to the discrete velocity method solving the Boltzmann-BGK equation (especially the linearized-BGK equation), which we will discuss in the section below. 
  
\subsection{Discrete velocity methods and lattice Boltzmann method}

The above procedure of establishing LB models is similar to the
problem solving process of the discrete velocity method, which directly solves the
Boltzmann-BGK equation. Since DVM has been proved to be able to
provide accurate results for rarefied gas dynamics \cite[see][and
references therein] {Mieussens200183,MIEUSSENS20001,Mieussens2000,Yang1995323,PhysRevE.65.026315,aoki:2260,valougeorgis:521,naris:097106,2005PhFl...17j0607N,Sharipov2009,Sharipov2008}, it is helpful to compare two numerical methods in depth.

The discrete velocity method is to discretize the velocity space based on quadratures e.g. Gauss-Hermite quadrature and Newton-Cotes quadrature \cite[see][]{2005PhFl...17j0607N,
  naris:097106,valougeorgis:521,Yang1995323}.  The first step is to non-dimensionalize the Boltzmann-BGK equation and obtain the reduced functions, e.g. $\mathcal{G}_a$ and $\mathcal{G}_b$ in Eqs.(\ref{ga}) and (\ref{gb}), which are important to reduce computational
costs.  The second step is to apply an appropriate discretization method
for the velocity space, which
is important but difficult because the velocity space ranges from $-\infty$
to $+\infty$ and the properties of conservation and dissipation of the
entropy should be kept. A typical choice is the Gauss-Hermite quadrature, which is to be
adopted in our simulations. In order to reduce the velocity components which
need to  be integrated from  $-\infty$ to $+\infty$, curvilinear coordinates including the polar
coordinates for 2D systems may be used for the velocity space. Afterwards,
the continuous Maxwell equilibrium should also be discretized. The last step is
to adopt an appropriate numerical scheme for the space and time
discretization.  Therefore, we can see that
LB methodology closely resembles the DVM problem solving
process. \cite{Luo200063} noticed this similarity and stated \textbf{``}the LB equation is essentially DVM with finite
discrete velocities and fully discretized space and time tied to the discrete
velocity set\textbf{"}. For simulating rarefied gas flows, this similarity is important as we
have shown how the LB framework is developed from the Boltzmann-BGK equation.

For both DVM and LB methods, the most critical task is to discretize the
velocity space. When the Gauss-Hermite quadrature is used in DVM, the
discretization of the velocity space in these two methods are the same, which
may indicate that the LB models with sufficiently accurate Gauss-Hermite
quadrature can capture the higher-order non-equilibrium effects in the rarefied
gas flows.  This in deed is confirmed by the simulation results
 presented in Fig.\ref{exm}, which we will discuss in detail in Section 4. 

However, an important advantage of the LB models is the ``stream-collision'' mechanism
which is mainly inherited from the lattice gas automata. This
``stream-collision'' mechanism makes the LB method easy to understand and simple for computer programming. Therefore, the ``stream-collision'' mechanism is an
important feature of the LB models which distinguishes them from DVM. The coupled time step and physical space in the LB models will
dramatically reduce the computational cost.  In addition, DVM relies heavily on mathematical techniques
which depend on specific problem, while the LB methodology is straightforward and more suitable for developing a generic
simulation package for engineering design.

\subsection{Lattice Boltzmann equation and linearized BGK equation}

By introducing $\psi$ to denote the unknown perturbed distribution function and
assuming the flow is weakly non-equilibrium, $f$ can be approximated by
\begin{equation}
\label{lsum}
  f=f^0(1+\psi),
\end{equation}
where
\begin{equation}
  f^0=\frac{1}{(2\pi)^{D/2}} \text{e}^{-\xi^2/2},
\end{equation}
which is the global (absolute) equilibrium distribution function. Using the
Taylor series to expand the local equilibrium distribution function and keeping
the terms up to the first order, one can obtain the following equation

\begin{equation}
\label{generalbgk}
\frac{\partial f}{\partial t} + \bm \xi \cdot \nabla f + \bm g \cdot \nabla_\xi f=
-\frac{1}{Kn}\left\{f-f^0\left[1 + \bm \xi \cdot \bm u + \frac{1}{2}(T-1)(\xi^2-D)\right] \right\},
\end{equation}
where we assume the flow is incompressible. Using Eq.(\ref{lsum}),
we can obtain the linearized BGK equation:
\begin{equation}
\label{generallbgk}
\frac{\partial \psi}{\partial t} + \bm \xi \cdot \nabla \psi + \bm g \cdot
\left[\nabla_\xi  \psi -\left( 1+\psi)\bm \xi \right)\right]=
-\frac{1}{Kn}\left\{\psi-\left[\bm \xi \cdot \bm u + \frac{1}{2}(T-1)(\xi^2-D)\right] \right\}.
\end{equation}
For lattice Boltzmann models, one can rewrite Eq.(\ref{approxf}) as
\begin{equation}
\label{approxfhermite}
f(\bm r, \bm \xi,t) \approx f^{N}(\bm r, \bm \xi,t)=\omega(\bm \xi)\left[1 +\varphi(\bm
r,\bm \xi,t) \right],
\end{equation}
where 
\begin{equation}
\varphi(\bm r, \bm \xi,t)=\sum^N_{n=1}\frac{1}{n!}\bm a^{(n)}(\bm r,t) \bm \chi^{(n)}(\bm \xi).
\end{equation}
Substituting Eq.(\ref{approxfhermite}) into the Boltzmann-BGK equation and
keeping the first- and second-order expansions of the equilibrium distribution, one can obtain

\begin{equation}
\label{lbm1}
\frac{\partial \varphi}{\partial t} + \bm \xi \cdot \nabla \varphi + \bm g \cdot
\left[\nabla_\xi  \varphi -\left( 1+\varphi\right) \bm \xi \right]=
-\frac{1}{Kn}\left (\varphi-\bm \xi \cdot \bm u \right),
\end{equation}
and
\begin{equation}
\label{lbm2}
\frac{\partial \varphi}{\partial t} + \bm \xi \cdot \nabla \varphi + \bm g \cdot
\left[\nabla_\xi  \varphi -\left( 1+\varphi \right) \bm \xi  \right]=
-\frac{1}{Kn}\left \{\varphi-\bm \xi \cdot \bm u -
    \frac{1}{2}\left[(\bm \xi \cdot \bm u)^2-u^2+(T-1)(\xi^2-D)\right] \right\}.
\end{equation}

Because $\omega(\xi)$ is equal to $f_0$, we can observe the following interesting facts by comparing Eqs.(\ref{lbm1}) and (\ref{lbm2}) with  Eq.(\ref{generallbgk}). First of all, by keeping the first order Hermite expansion,
  the essential LB model equation is the same with that of the isothermal ($T=1$) linearized BGK equation except the body force term. {\it This implies that $\varphi$ is indeed equivalent to $\psi$ though $\varphi$
  is prescribed to include only the finite order terms of the Hermite polynomials
  (cf. Eq.(\ref{approxf}) and Eq.(\ref{approxfhermite})). Therefore, the LB equation with the first order terms should be as good as the linearized BGK equation for isothermal flows}. This indicates that high-order Hermite expansion is not necessary for rarefied gas flows. Secondly, with the second order Hermite
  expansion, there is an extra velocity term $ \frac{1}{2}\left[(\bm \xi \cdot
    \bm u)^2-u^2\right] $ for the LB equation in comparison to the linearized BGK equation. However, for flows with low Mach number, this term is a higher-order small quantity which can  be ignored. This is the
  reason why the Hermite expansion order is reported to make negligible
  difference on the simulation results \cite[see][]{Kim20088655}. In fact, the first
  order expansion is sufficient to obtain the accurate results for isothermal rarefied flows with low speed. Furthermore, the LB
  equation with the second order expansion can in principle describe thermal problems since the
  temperature information is included in Eq.(\ref{lbm2}), which at
  least has the same capability as the linearized BGK equation, though the BGK kinetic
  model gives wrong $Pr$ number.  Thirdly, the treatment of the body force makes
  the difference between the LB model and the linearized BGK equation. It is because that the linearized BGK model keeps
  the full information while the LB model uses the Hermite expansion to
  approximate $\nabla_\xi f$, i.e. $\nabla_\xi  \varphi -\left( 1+\varphi \right)\bm
  \xi$. However, for the problem is not far from equilibrium state, this
  difference is not important, which will be confirmed by the numerical simulations in Section 4. 

  From the above analysis, we can see that the Hermite expansion order does not determine the accuracy of LB
  models for rarefied gas flows as described by Eq.(4.7) in
  \cite{{2006JFM...550..413S}}. The Hermite expansion provides a means to approximate the equilibrium distribution
  and the body force in the kinetic equation. Therefore, the LB equation, similar to the linearized BGK equation, is an approximation of the Boltzmann-BGK equation. In contrast to the Grad's moment method, LB models include the
  information of any order moment though it may not be accurate. For instance,
  with the first order expansion, the LB model equation is as the same as the isothermal
  linearized BGK equation in the incompressible limit, which will give
  accurate results for any order velocity moment. When the Mach number of flow increases, high-order terms in the Hermite expansion become important. Therefore, the order of Hermite expansion is important to simulate compressible flows rather than rarefied flows.

  To capture non-equilibrium effects in rarefied flows, the Gauss-Hermite quadrature is the key as it determines
  the discretization accuracy to the model equation. With sufficiently high order of the Gauss-Hermite quadrature, LB models can give excellent
  numerical results, e.g. the results presented in Fig.(\ref{exm}) where 400 discrete velocities are used are identical to the DVM solution. Considering the similarity of the LB equation and linearized BGK equation, insufficient quadrature order should be responsible for the failure
  on capturing the constitutive relations in the Knudsen layer because the kinetic boundary condition have been well accepted in solving the linearized BGK equation.

In summary, the LB method is essentially a special discrete velocity
model, which approximates the Boltzmann-BGK equation with finite discrete
velocities and fully discretized space and time tied to the discrete velocity
set. The capability of LB equation is similar to the linearized BGK equation for simulating rarefied gas flows. The Hermite expansion order determines the model equation and is important for compressible flows. It has no direct effect on the accuracy of capturing high-order non-equilibrium effects. Meanwhile, the Gauss-Hermite quadrature as a discretization technique for the velocity space directly determines whether the LB models can describe rarefied flows accurately.  
       
\section{Simulations and discussion}

   
In addition to the above theoretical analysis, we will numerically evaluate the LB models. To exclude the boundary condition effect, we choose the standing-shear-wave problem as the benchmark case, which was specially designed for
assessing the accuracy of various models \cite[][]{Lockerby2008}. It is a shear flow
driven by a temporally and spatially oscillating body force, which can be written
as the following form:
\begin{equation}
  \label{force}
  F_x=A \text{e}^{\text{i} \phi t} \cos \theta y,
\end{equation}
where $F_x$ is the body force in the direction $x$ (which is perpendicular to
the $y$ direction), $A$ is the amplitude, and $\theta$ is the wave number and
$\phi$ is the frequency. This isothermal problem is sufficiently simple because
the flow direction is perpendicular to the space variation but it is intended to
capture the shear-dominated characteristic of microscale flows. Furthermore, the
distinct advantage is that the boundary is not important here so that one can
focus on the model itself without the interference from gas molecule/wall interactions. With
Eq.(\ref{nondime}), the body force becomes:
\begin{equation}
  F_x=\hat{A} \text{e}^{\text{i} \phi t} \cos y,
\end{equation}
where $\theta$ is considered as a measure of the characteristic length, and $\hat{A}= \frac{\bm A}{\theta R T}$.  Another distinctive
advantage for using this benchmark problem is that analytical solutions can be
obtained for many hydrodynamic models, such as the Navier Stokes equation and
the regularized 13-moment model (R13). For convenience, the R13 solution is
listed as below:

\begin{equation}
\bar{u}=-\frac{\left(288{Kn}^{4}\text{i}-510\phi{Kn}^{3}+\left( 520\text{i}-225{\phi}^{2}\text{i}\right) {Kn}^{2}-375\phi Kn+150\text{i}\right) \hat{A}}{288 \phi {Kn}^{4}+\left( 510 {\phi}^{2} \text{i}-270 \text{i}\right) {Kn}^{3}+\left( 745 \phi-225 {\phi}^{3}\right) {Kn}^{2}+\left( 375  {\phi}^{2} \text{i}-150 \text{i}\right) Kn+150 \phi}, 
\end{equation}
where $\bar{u}$ denotes the velocity amplitude. One can refer to
\cite{Lockerby2008} for the detail of hydrodynamics models.   

The discrete velocity method of solving the linearized BGK equation has already been served as a benchmark for
the standing shear wave problem by \cite{Lockerby2008}, where the linearized BGK model Eq.(\ref{generallbgk}) can be simply written in the scalar form for isothermal flows:

\begin{equation}
  \label{LBGK}
  \frac{\partial \psi}{\partial t} + \xi_y \frac{\partial \psi}{\partial y} + F_x \left[ \frac{\partial \psi}{\partial \xi_x}-(1+\psi)\xi_x\right]=
  \frac{1}{Kn}\left(\xi_x u_x - \psi \right).
\end{equation}
Since the problem is essentially one-dimensional, one can eliminate
$\xi_x$ by multiplying the above equation with
$\frac{1}{\sqrt{2\pi}}\text{e}^{-\xi_x^2/2}$ and
$\frac{1}{\sqrt{2\pi}}\xi_x\text{e}^{-\xi_x^2/2}$ respectively. Integrating over
$\xi_x$, the resulting equations are:

\begin{equation}
  \label{ga}
  \frac{\partial \mathcal{G}_a}{\partial t}+ \xi_y\frac{\partial \mathcal{G}_a}{\partial y}= -\frac{1}{Kn}\mathcal{G}_a,
\end{equation}

\begin{equation}
  \label{gb}
  \frac{\partial \mathcal{G}_b}{\partial t}+ \xi_y\frac{\partial \mathcal{G}_b}{\partial y} -F_x(\mathcal{G}_a+1) = \frac{1}{Kn}(u_x-\mathcal{G}_b),
\end{equation}
where the reduced unknown functions $\mathcal{G}_a$ and $\mathcal{G}_b$ are
defined as
\begin{equation}
  \mathcal{G}_a=\frac{1}{\sqrt{2\pi}}\int^{\infty}_{-\infty}\psi \text{e}^{-\xi_x^2/2}d \xi_x,
\end{equation}
\begin{equation}
  \mathcal{G}_b=\frac{1}{\sqrt{2\pi}}\int^{\infty}_{-\infty}\xi_x \psi \text{e}^{-\xi_x^2/2} d \xi_x.
\end{equation}
The macroscopic velocity can be expressed as
\begin{equation}
  u_x=\frac{1}{\sqrt{2 \pi}} \int^{\infty}_{-\infty}\mathcal{G}_b d \xi_y.
\end{equation}

To solve Eqs.(\ref{ga}) and (\ref{gb}), the essential task is to choose an
appropriate quadrature to discretize the velocity space which ranges from
$-\infty$ to $\infty$. The typical highly accurate choice for low speed
rarefied gas flows is the Gauss-Hermite quadrature, which is used here. Based on the
discretization  of the phase
space, the integration operation over the velocity space is converted to sum
operation and then a series of equations like Eq.(\ref{lbgk}) are
obtained. Naturally, the discretized Maxwell equilibrium distribution can also be obtained by directly using its value on the grid of the velocity
space. One can then use typical numerical methods such as finite difference
scheme (e.g., the Lax-Wendroff scheme) to solve these equations respectively.

\begin{figure}
  \centering
  \includegraphics[height=7cm,width=12cm]{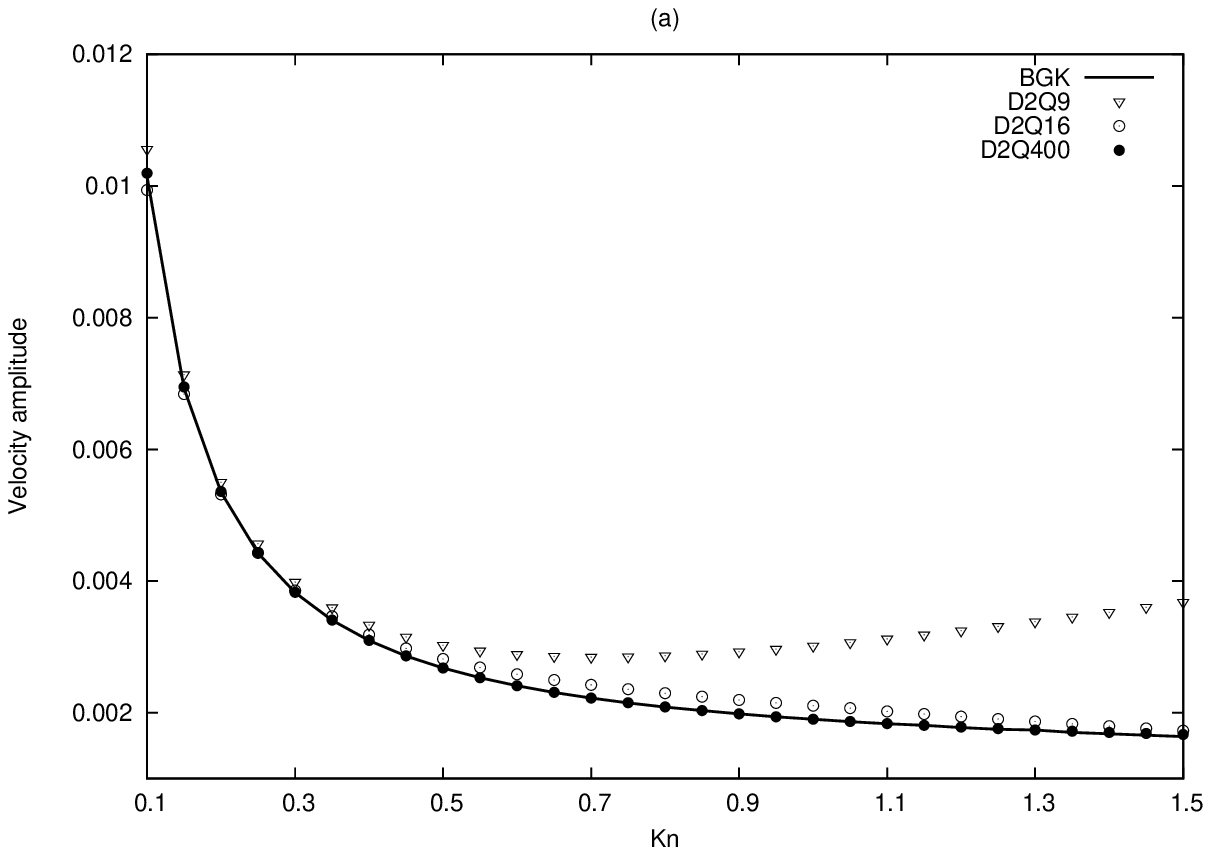}
  \includegraphics[height=7cm,width=12cm]{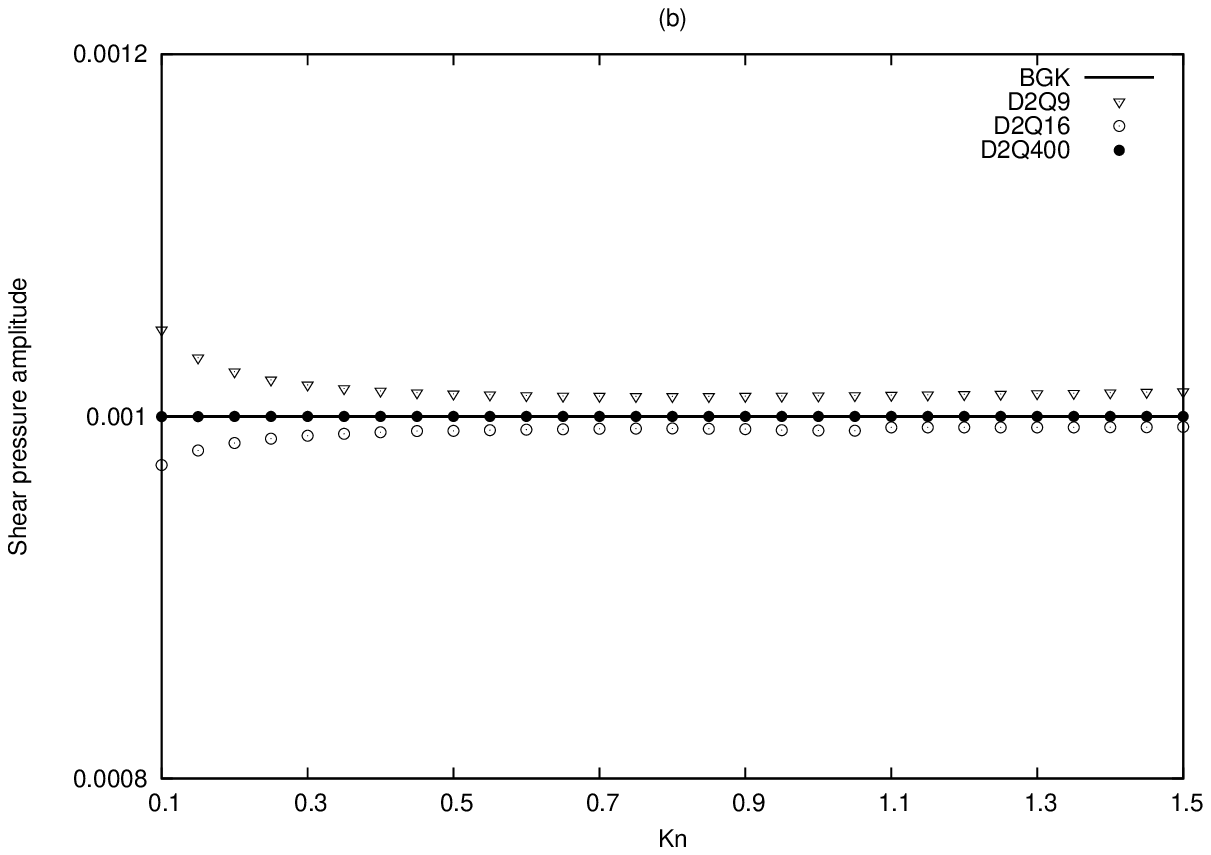}
  \caption{The results of D2Q400, D2Q16 and D2Q9 models for the quasi-steady standing shear wave (a) velocity
    wave amplitude,(b) shear pressure wave amplitude. The first-order Hermite expansion is adopted for the D2Q400 model. Since the Hermite
    polynomials for the D2Q400 model give irrational roots, the Lax-Wendroff scheme is used to solve
    Eq.(\ref{lbgk}) here. The results show that the LB model with sufficiently
    large discrete velocity sets can be very accurate.}
  \label{exm}
\end{figure}

When the same Gauss-Hermite quadrature with 400 discrete velocities are used in the DVM
solution of the linearized BGK equation and our LB model, Fig.\ref{exm} shows that the results for both velocity and shear pressure amplitude are nearly identical for a broad range of Knudsen numbers from 0.1 to
1.5. Even with the first order Hermite expansion, the LB model can predict shear pressure accurately, which confirms that the Hermite expansion order does not directly affect accuracy of the LB models in capturing non-equilibrium effects measured by the Knudsen number.    

Although the standard LB model (D2Q9) is not sufficiently accurate in
comparison  with the DVM solution, high-order LB model (D2Q16) with minimal
increase  of the discrete velocity set can produce good
results. Fig.\ref{exm} shows that the LB model with increasing order
of the Gauss-Hermite quadrature can closely approximate the linearized BGK
equation. Therefore, in comparison with the DVM simulation, LB method can provide a practical engineering design simulation tool which can produce reasonably accurate results with significantly reduced computational cost. 

As discussed in Section 2.3, at least three factors will influence the problem-solving
process, i.e. the numerical scheme for solving Eq.(\ref{lbgk}), the order
of Hermite expansion and Gauss-Hermite quadrature. For the numerical scheme,
our second-order scheme is essential as discussed in Section 2.2. Regarding the
role of Hermite expansion and Gauss-Hermite quadrature, we have theoretically proved that the Gauss-Hermite quadrature rather than the order of Hermite expansion is key to capturing non-equilibrium effects accurately. The numerical simulations have also performed to testify our theoretical analysis. 

\begin{figure}
  \centering
  \includegraphics[height=7cm,width=12cm]{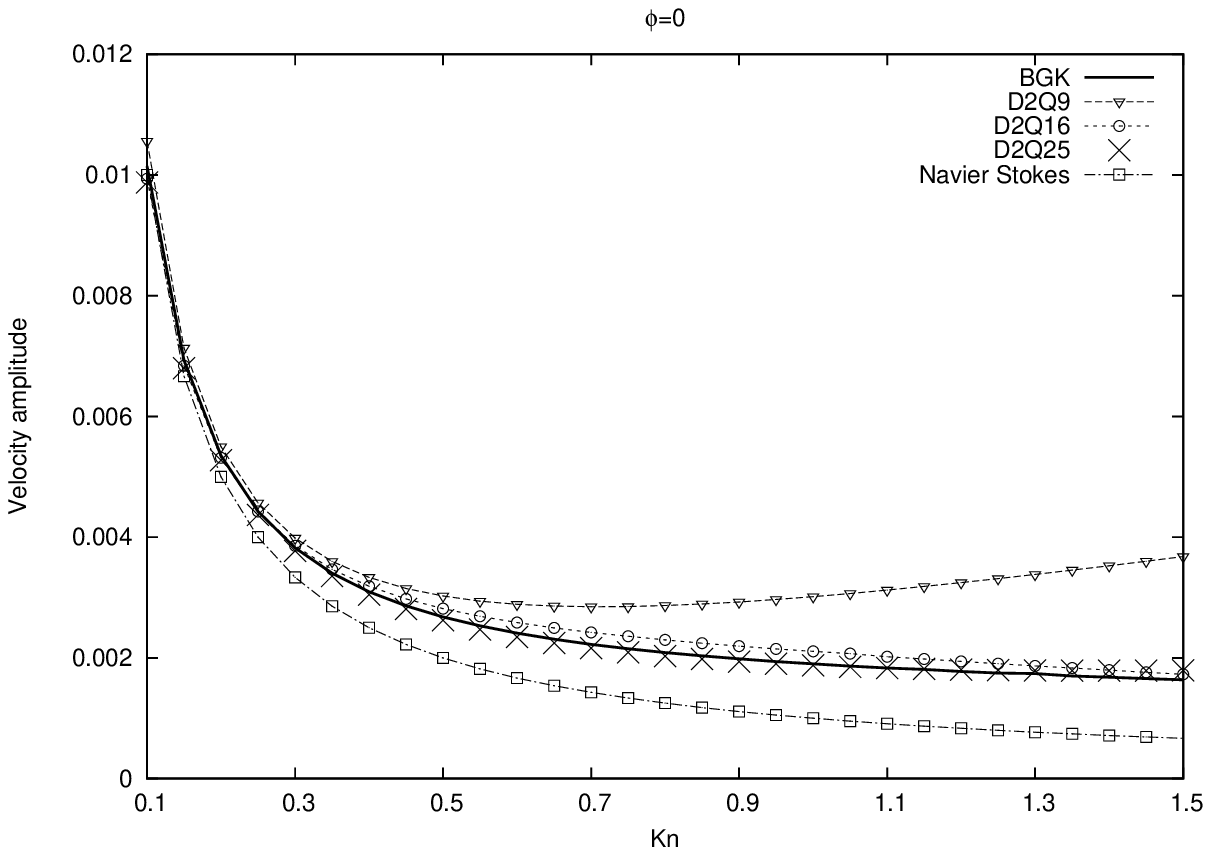}\\
  \includegraphics[height=7cm,width=12cm]{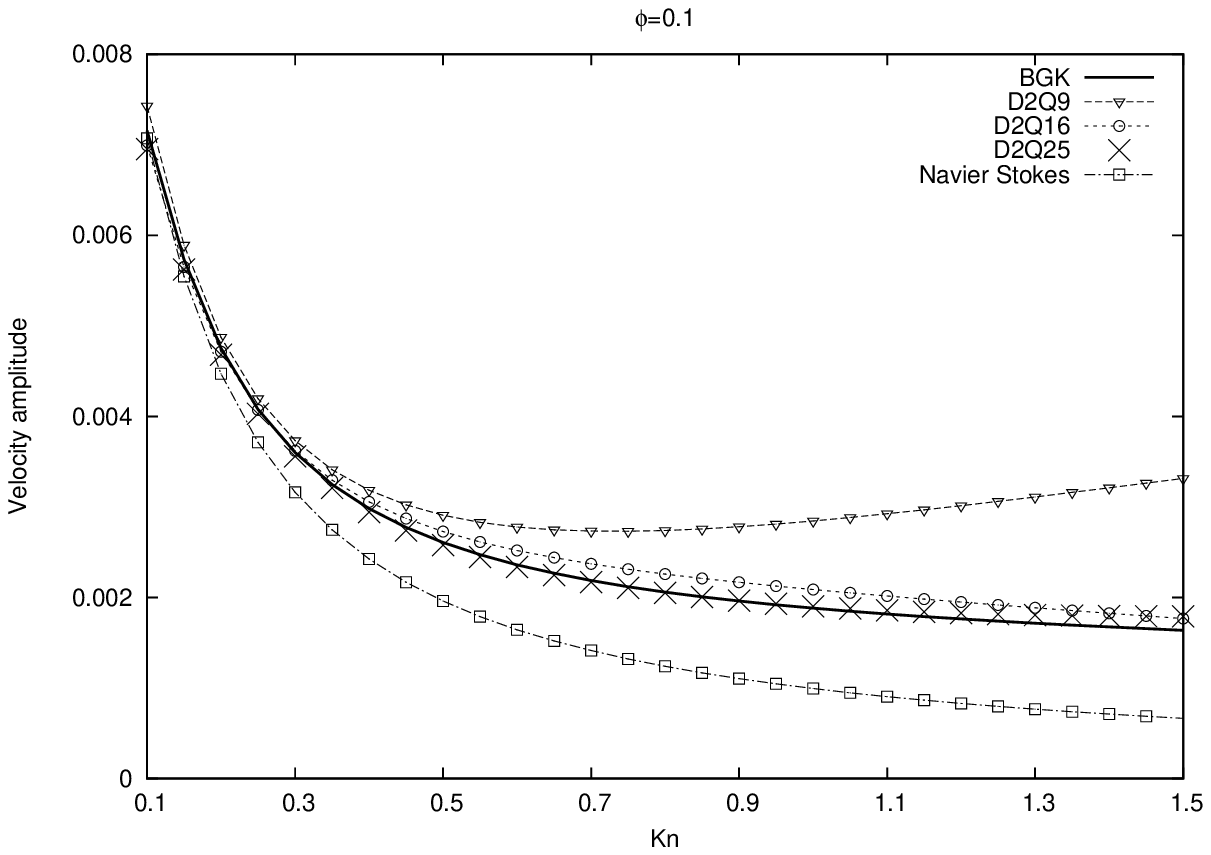}\\
  \includegraphics[height=7cm,width=12cm]{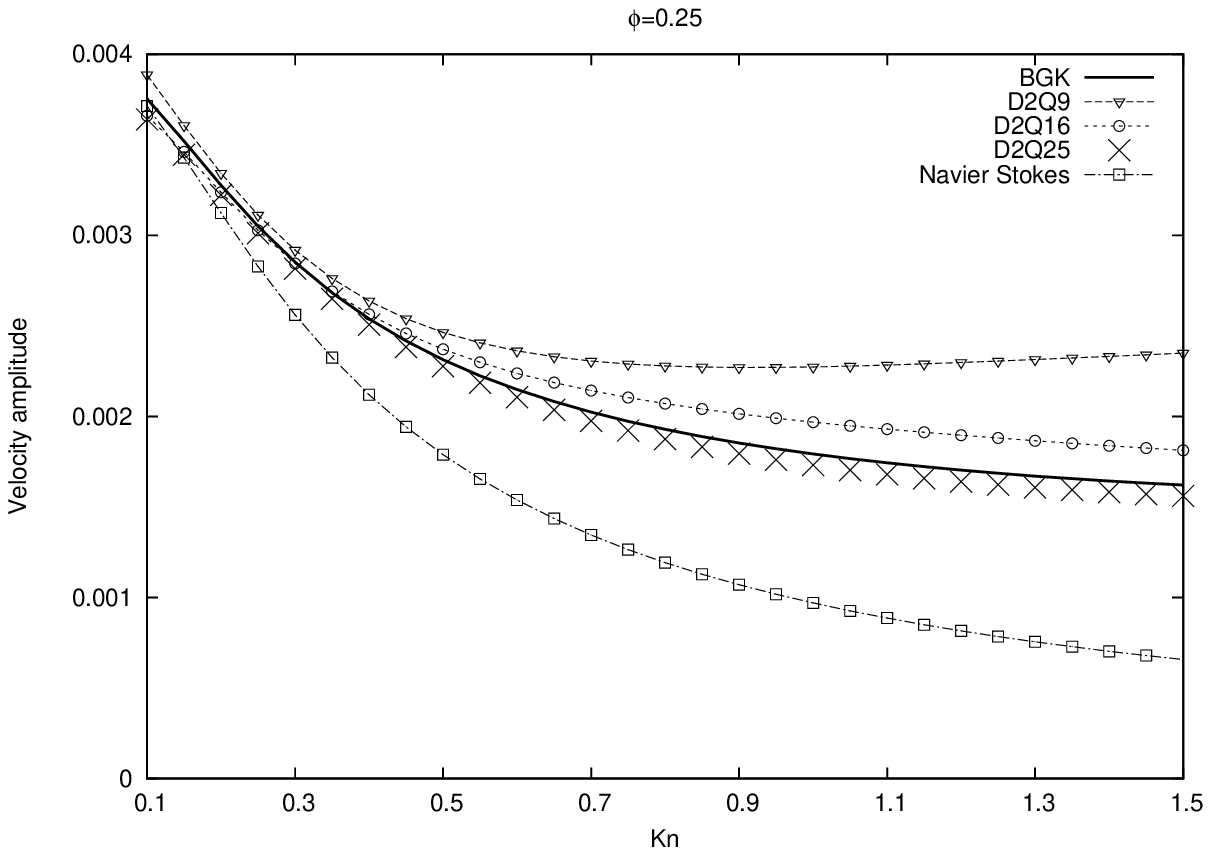}
  \caption{Velocity wave-amplitude as a function of the Knudsen number, where the expansion order for the equilibrium distribution function and the force term is $N$ which is 2, 3, 4 for the D2Q9, D2Q16 and D2Q25 models respectively, and the order of Gauss-Hermite quadrature is $2N+1$.}
  \label{normal}
\end{figure}

\begin{figure}
  \centering
  \includegraphics[height=7cm,width=12cm]{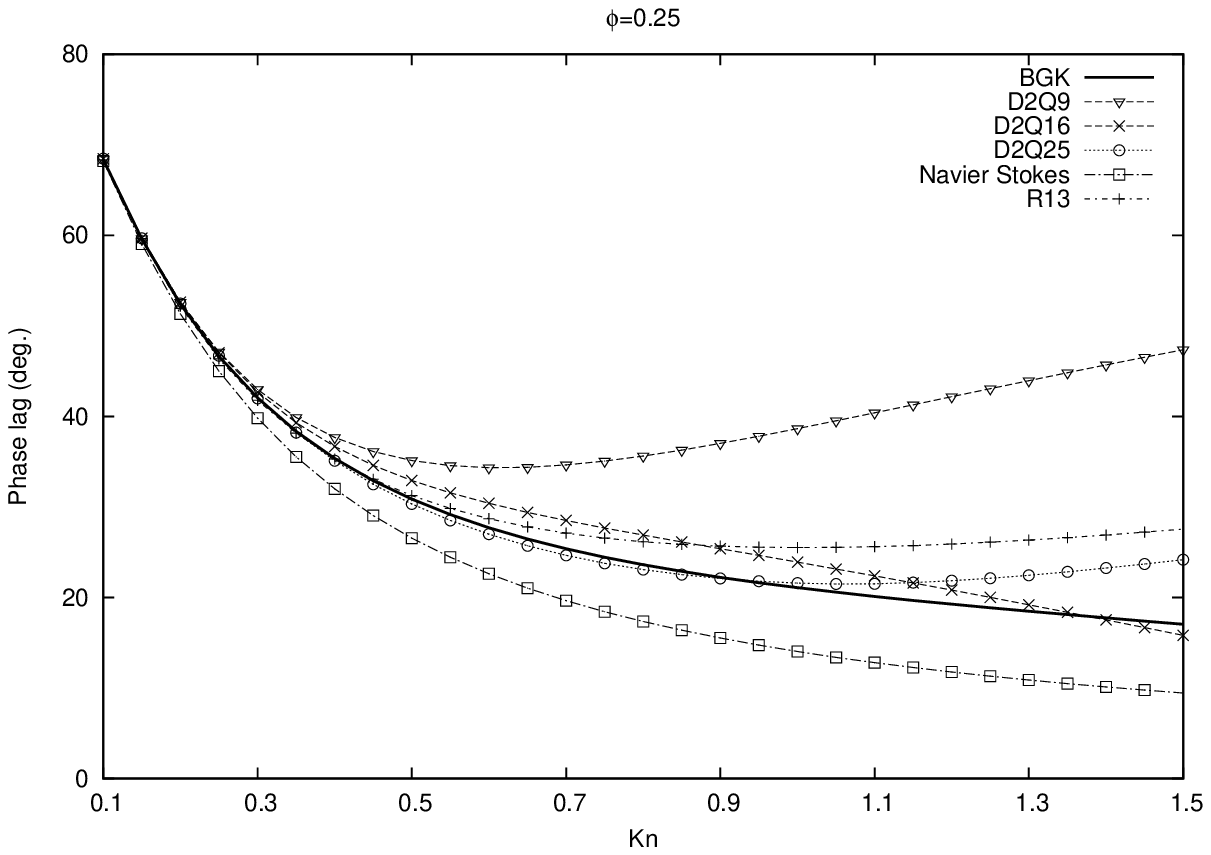}\\
  \caption{Velocity wave phase lag as a function of the Knudsen number, where the expansion order for the equilibrium distribution function and the force term is $N$ which is 2, 3, 4 for the D2Q9, D2Q16 and D2Q25 models respectively, and the order of Gauss-Hermite quadrature is $2N+1$.}
  \label{lag}
\end{figure}

In Figs.\ref{normal} and \ref{lag}, the simulation results of the three LB
models are compared with the solutions of directly solving the linearized BGK
equation and the Navier Stokes equation. The expansion of the equilibrium
distribution function and the forcing term is second-order for the D2Q9 model,
third-order for the D2Q16 and forth-order for D2Q25. The results in Fig.\ref{normal} show that the
prediction for velocity amplitude of the D2Q25 model are in excellent agreement
with the DVM solution of the linearized BGK equation across a broad range of Knudsen number ($Kn\in[0,1.5]$) for the quasi-steady and time-varying problems with $\theta$ up to $0.25$. Meanwhile, the results of the D2Q9 model  deviate from the DVM
solution of the linearized BGK equation significantly. Surprisingly, the D2Q9 model does not
agree with the results predicted by the Navier Stokes equation. Fig.\ref{lag} shows the velocity wave phase
lag, which suggests that high-order LB models perform better in the transition flow regime.  

\begin{figure}
  \centering
  \includegraphics[height=7cm,width=12cm]{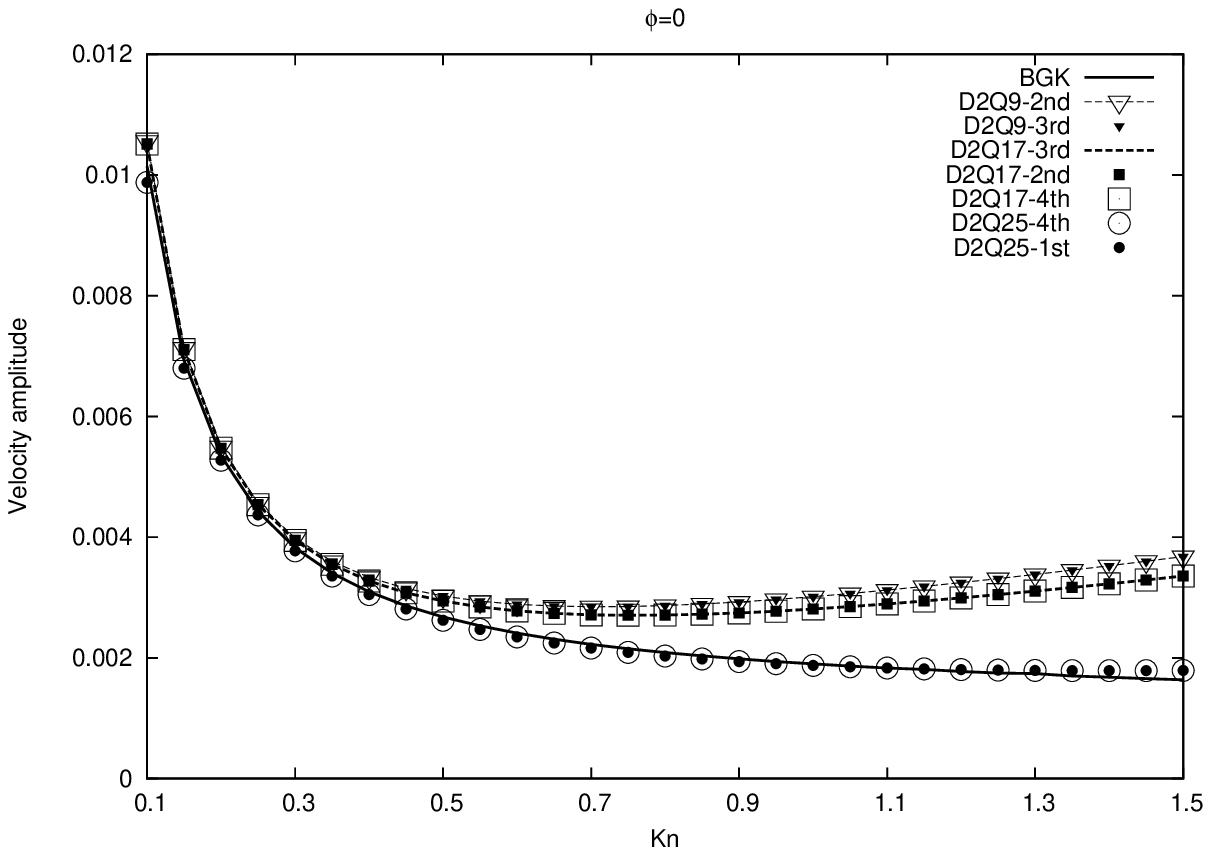}\\
  \includegraphics[height=7cm,width=12cm]{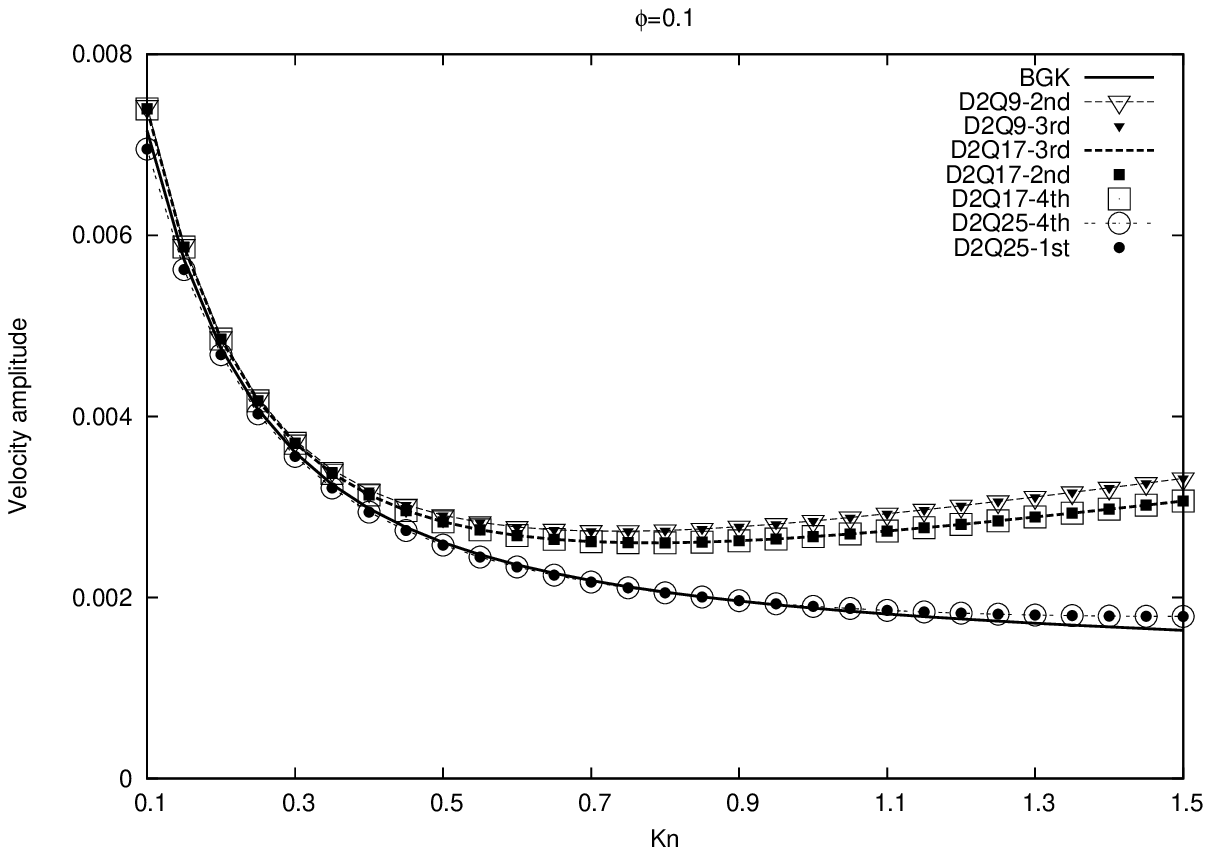}\\
  \includegraphics[height=7cm,width=12cm]{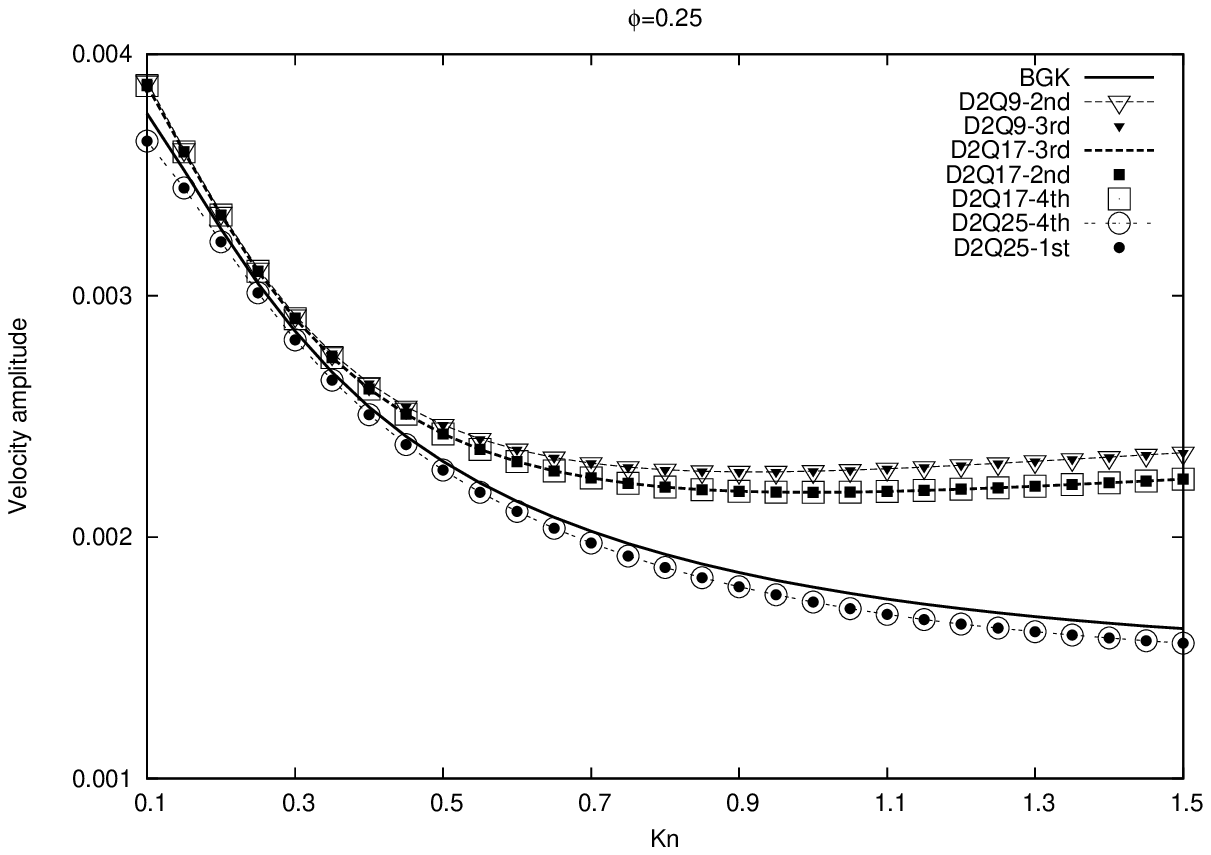}
  \caption{Velocity wave-amplitude varying with the Knudsen number. The models are
    named according to the rule D2Qn - Yth where n denotes the number
    of discrete velocities, Y the expansion order for the
    equilibrium expansion and the force term.}
  \label{diff}
\end{figure}

Although Figs.\ref{normal} and \ref{lag} demonstrate that increasing order of LB model in terms of the Hermite expansion and Gauss-Hermite quadrature will lead to more accurate results, we still do not know the exact role the orders of
the Hermite expansion and the Gauss-Hermite quadrature play.  Therefore, we
single out the effect of the Hermite expansion in Fig.\ref{diff}, where the results of the LB models with the same quadrature but
different Hermite expansion order are compared. The results clearly show that
the Hermite expansion order for both the force and the equilibrium distribution
function make negligible difference to the simulation results. Even the first order
expansion is sufficient to obtain accurate velocity for the D2Q25 model. The simulation results support our
theoretical analysis that the Hermite expansion has no direct influence on model accuracy for capturing non-equilibrium effects. Specifically, the LB model equation determined by the first order Hermite expansion is sufficient for a typical gas flow in micro-devices where the Mach number is usually small. In contrast, the Gauss-Hermite quadrature determines the model accuracy as the higher-order quadratures give better results.

\begin{figure}
  \centering
  \includegraphics[height=7cm,width=12cm]{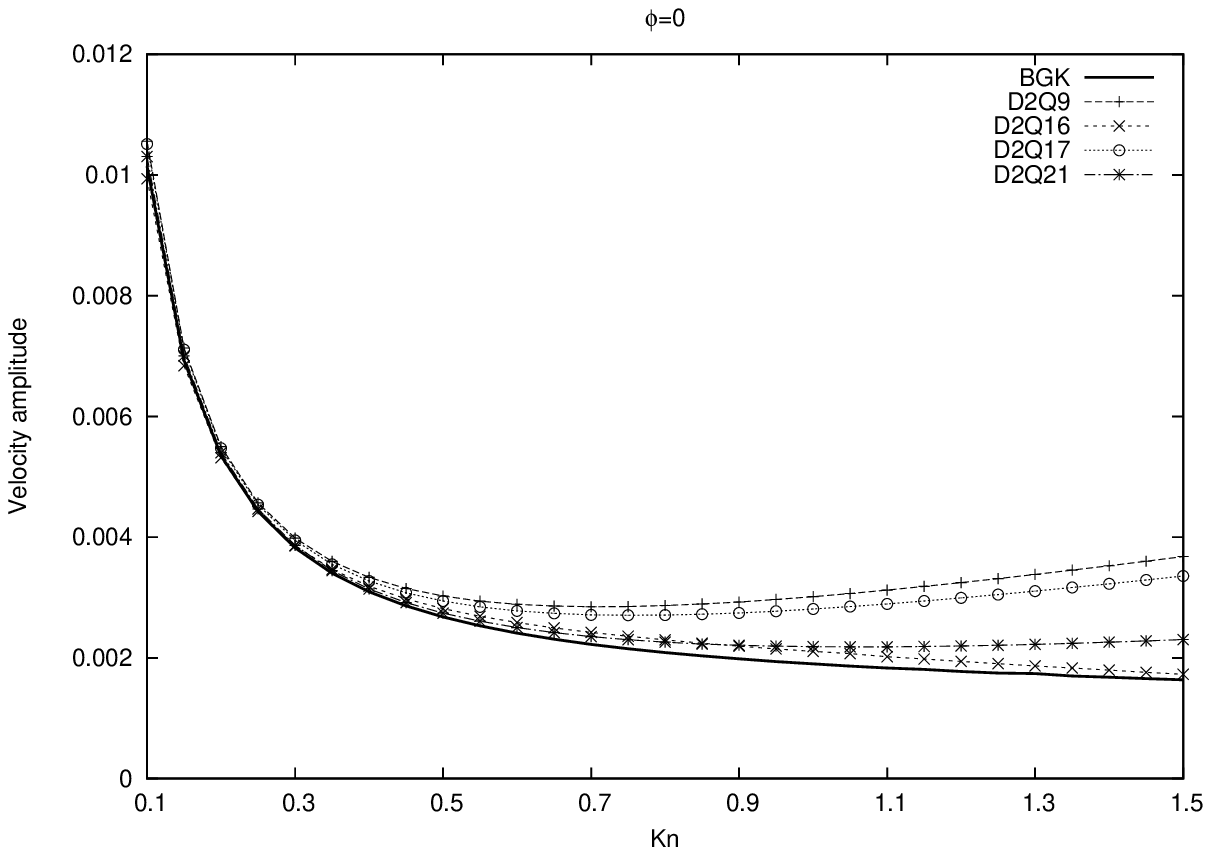}\\
  \includegraphics[height=7cm,width=12cm]{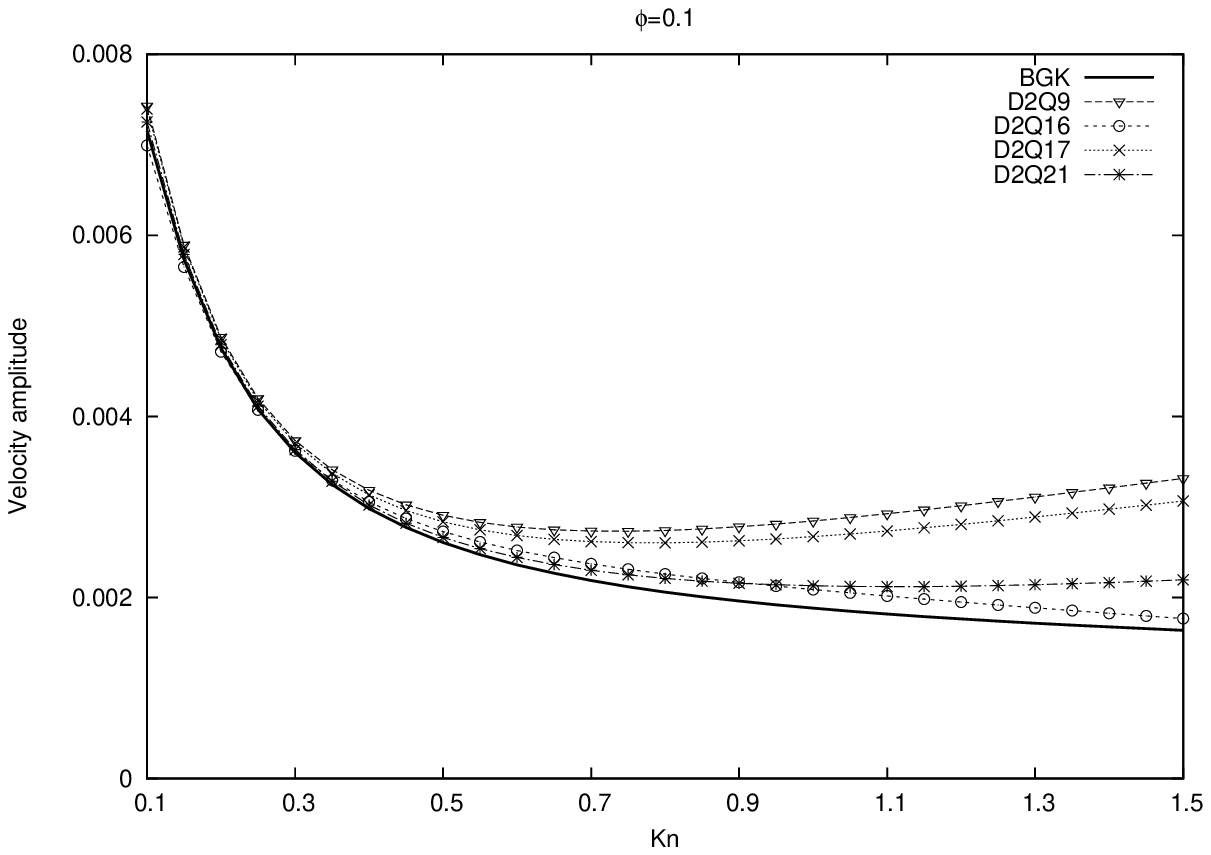}\\
  \includegraphics[height=7cm,width=12cm]{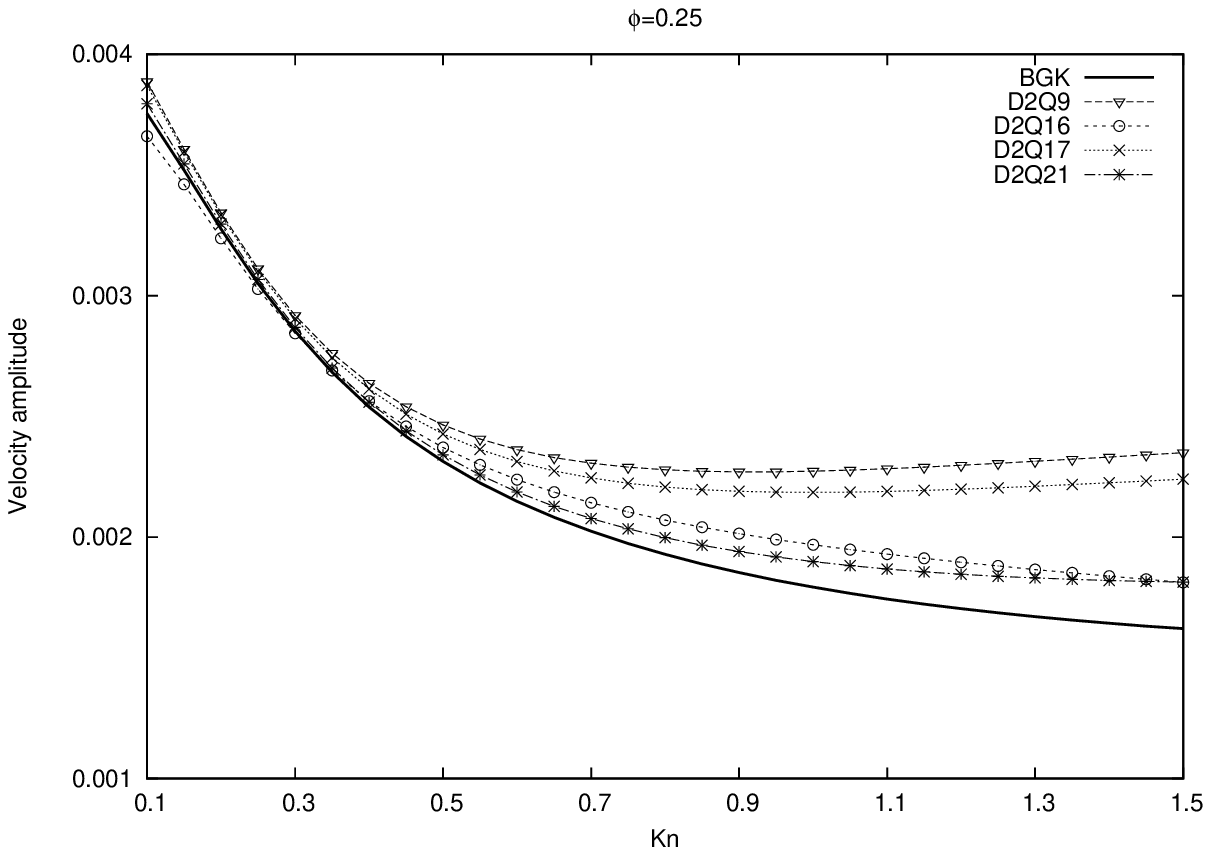}
  \caption{Velocity wave-amplitude varying with the Knudsen number, where the three models with the same order of quadratures but different abscissae are compared.}
  \label{three}
\end{figure}

Not only the order of quadrature but also the abscissae may influence the model
accuracy. Therefore, the simulation results of the three LB models with the same order quadrature but
different abscissae are compared in Fig.\ref{three}. Although increasing quadrature order will lead to improved
accuracy, more discrete velocities may not improve the model performance if the
quadratures are the same order. For example, the quadratures of the D2Q16, D2Q17 and D2Q21
models are the same order. Surprisingly, the D2Q16 model produces the results better than the other
two models with more discrete velocities. The reason may be attributed to that
the abscissae of the D2Q16 model has better symmetry. In addition, all these models
are better than the D2Q9 model which has low order quadrature. Therefore, appropriate abscissae may improve the model accuracy and reduce the computational costs with smaller number of discrete velocities.

\begin{figure}
  \centering

  \includegraphics[height=7cm,width=12cm]{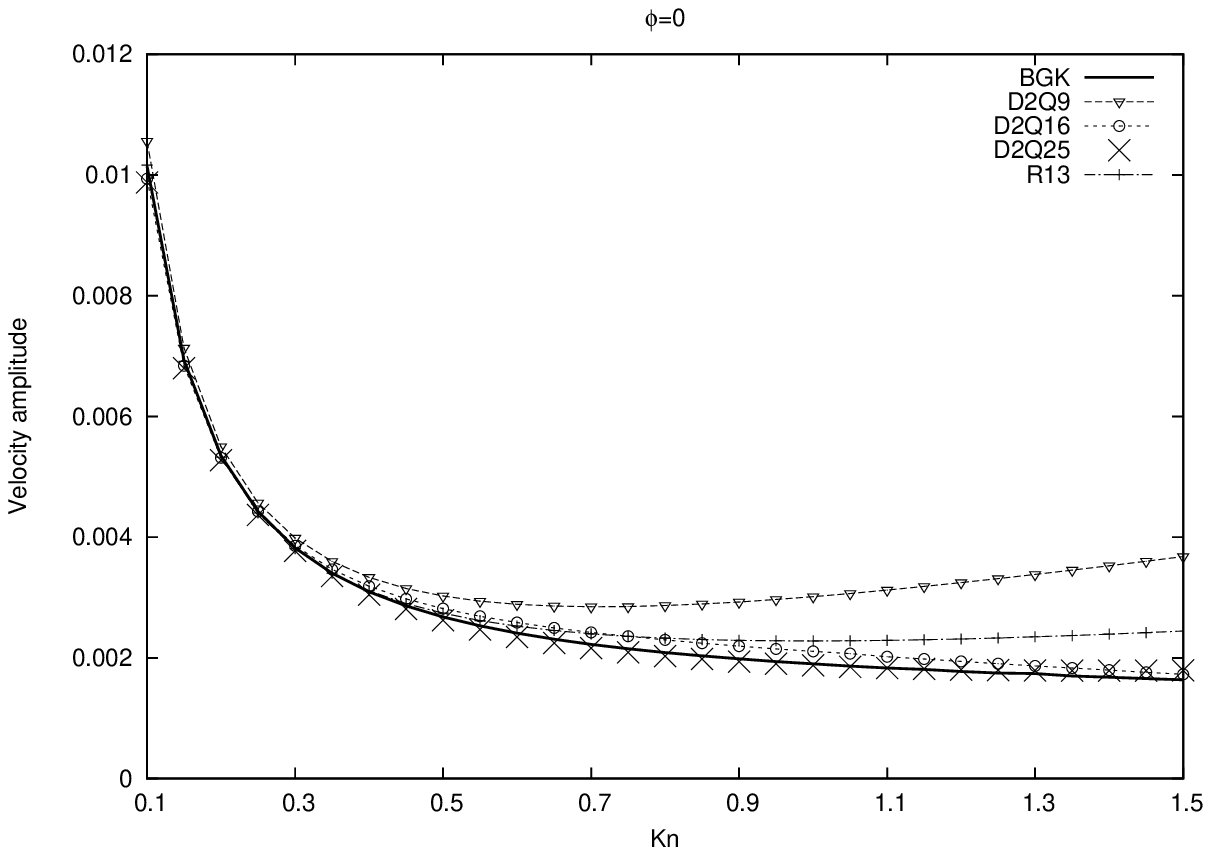}\\
  \includegraphics[height=7cm,width=12cm]{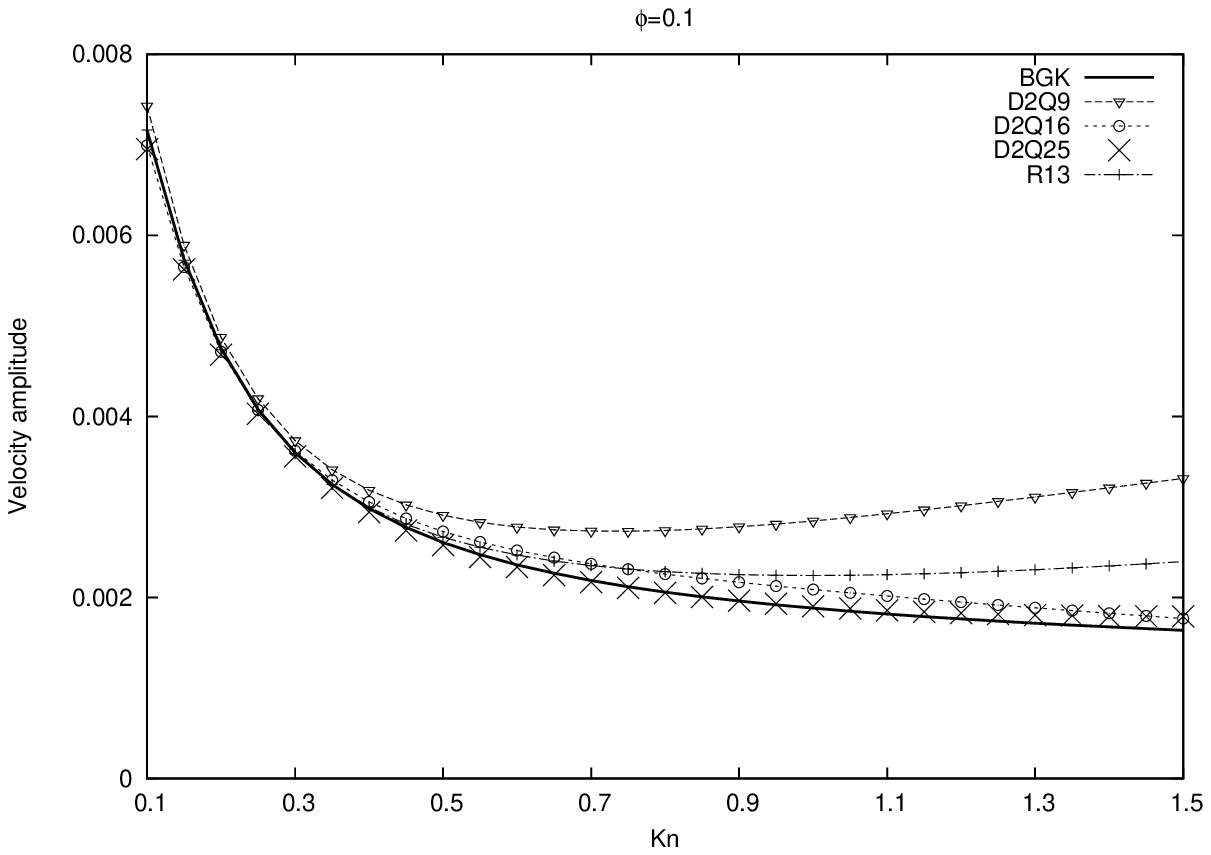}\\
  \includegraphics[height=7cm,width=12cm]{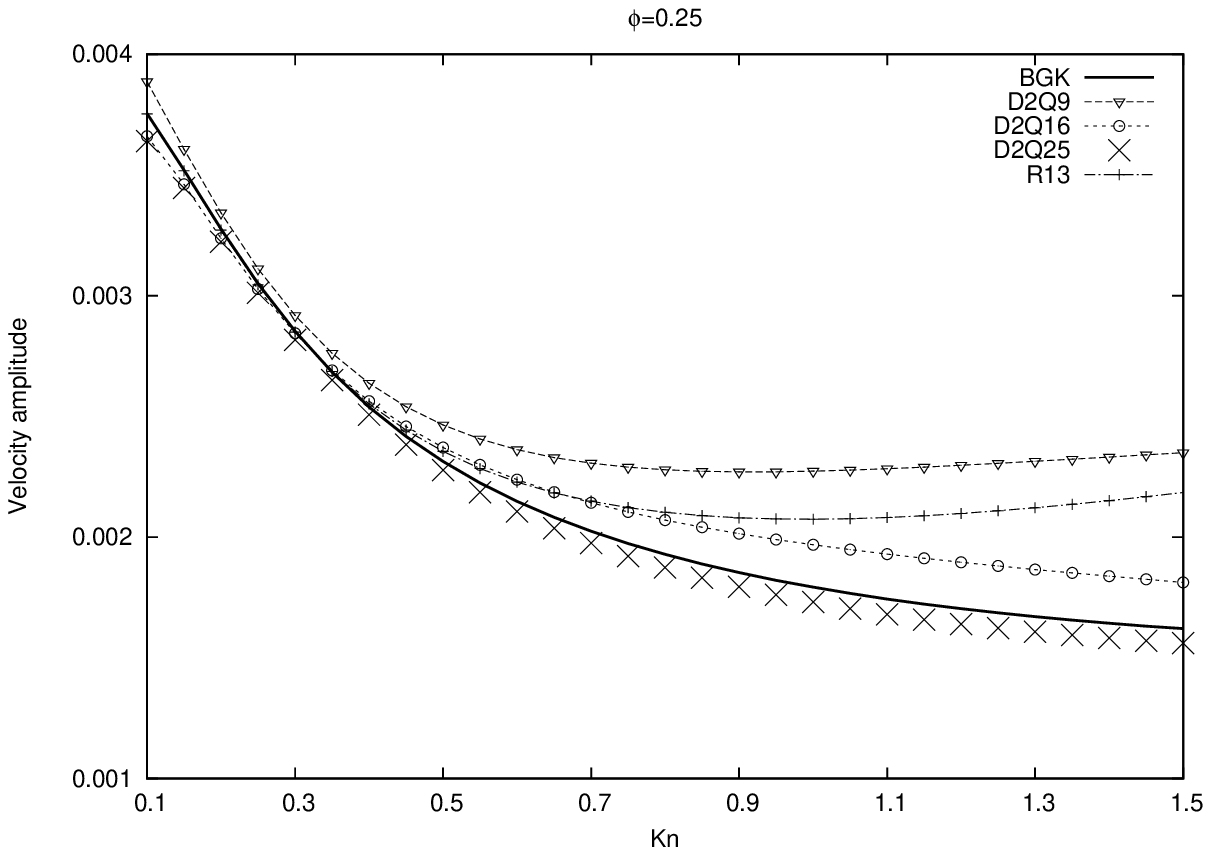}
  \caption{Velocity wave-amplitude varying with the Knudsen number where the results of
    LB models are compared against the solution of the R13 equation.}
  \label{r13e}
\end{figure}

Since \cite{Lockerby2008} has shown that the R13 equation gives the best
performance among the extended hydrodynamic models, we compare the LB
models with the R13 model here. Fig.\ref{r13e} shows that, in comparison with the data obtained from directly solving the linearized BGK
equation, the high-order LB models including the D2Q16 and D2Q25 models can give
better results than the R13 equation over a broad range of Knudsen numbers. Therefore, the
high-order LB models with modest number of discrete velocity set, such as the D2Q16 and
D2Q25 models, can offer close approximation to the linearized BGK equation. Most importantly, these high-order LB models
achieve such degree of accuracy at a fraction of computational costs associated
with directly solving the linearized BGK equation.

\section{Concluding remarks}
We have theoretically and numerically analyzed the high-order LB models for rarefied gas
flows. The lattice Boltzmann equation is shown to be equivalent to the linearized BGK equation in the incompressible limit. When the same Gauss-Hermite quadrature is used,
both LB and DVM simulations produce results in excellent agreement across a broad range of
the Knudsen numbers. This suggests the importance of the Gauss-Hermite quadrature and the great potential of the LB method for modeling rarefied gas flows. While the Gauss-Hermite quadrature is of the most importance to capturing non-equilibrium effects, the first-order Hermite expansion on the equilibrium distribution function is sufficient to obtain the correct moments for isothermal flows e.g. increasing the Hermite expansion order further will not improve the model
accuracy. For the same order Gauss-Hermite quadratures, the chosen abscissae will influence the model accuracy and more discrete
velocities may not lead to improved model accuracy.

Overall, we have demonstrated that LB method offers a computationally efficient approach to
solve the BGK equation. We can choose a suitable LB model to meet different
requirement on model accuracy and computational efficiency, which offers an ideal flexible engineering design simulation tool to be able to simulate flows in the continuum and transition regimes. 

\section{Acknowledgments}
The authors would like to thank Jason Reese, Xiaojun Gu and Guihua Tang for many informative
discussions. This work was financially supported by the
Engineering and Physical Sciences Research Council U.K. under Grants No. EP/D07455X/1 and No. EP/
F028865/1.

\clearpage

\begin{thebibliography}{42}
\expandafter\ifx\csname natexlab\endcsname\relax\def\natexlab#1{#1}\fi

\bibitem[Al-Ghoul \& Chan~Eu(2004)]{physreve.70.016301}
{\sc Al-Ghoul, M. \& Eu, B.~C.} 2004 Generalized hydrodynamics and
  microflows. {\em Phys. Rev. E\/} {\bf 70}~(1), 016301.

\bibitem[Ansumali {\em et~al.\/}(2007)Ansumali, Karlin, Arcidiacono, Abbas \&
  Prasianakis]{Ansumali2007}
{\sc Ansumali, S., Karlin, I.~V., Arcidiacono, S., Abbas, A. \& Prasianakis,
  N.~I.} 2007 Hydrodynamics beyond {N}avier-{S}tokes: {E}xact solution to the
  lattice {B}oltzmann hierarchy. {\em Phys. Rev. Lett.\/} {\bf 98}~(12),
  124502.

\bibitem[Aoki {\em et~al.\/}(1991)Aoki, Nishino, Sone \& Sugimoto]{aoki:2260}
{\sc Aoki, K., Nishino, K., Sone, Y. \& Sugimoto, H.} 1991
  Numerical analysis of steady flows of a gas condensing on or evaporating from
  its plane condensed phase on the basis of kinetic theory: Effect of gas
  motion along the condensed phase. {\em Phys. Fluids A\/} {\bf 3}~(9),
  2260--2275.

\bibitem[Aoki {\em et~al.\/}(2002)Aoki, Takata \&
  Nakanishi]{PhysRevE.65.026315}
{\sc Aoki, K., Takata, S. \& Nakanishi, T.} 2002 Poiseuille-type
  flow of a rarefied gas between two parallel plates driven by a uniform
  external force. {\em Phys. Rev. E\/} {\bf 65}~(2), 026315.

\bibitem[{Balakrishnan}(2004)]{2004jfm...503..201b}
{\sc {Balakrishnan}, R.} 2004 {An approach to entropy consistency in
  second-order hydrodynamic equations}. {\em J. Fluid Mech.\/} {\bf 503},
  201--245.

\bibitem[{Benzi} {\em et~al.\/}(1992){Benzi}, {Succi} \&
  {Vergassola}]{1992phr...222..145b}
{\sc {Benzi}, R., {Succi}, S. \& {Vergassola}, M.} 1992 {The lattice
  {B}oltzmann equation: {T}heory and applications}. {\em Phys. Rep.\/} {\bf
  222}, 145--197.

\bibitem[{Chapman} \& {Cowling}(1991)]{1991mtnu.book.....c}
{\sc {Chapman}, S. \& {Cowling}, T.~G.} 1991 {\em {The Mathematical Theory of
  Non-uniform Gases}\/}. Cambridge University Press.

\bibitem[{Chen} \& {Doolen}(1998)]{1998anrfm..30..329c}
{\sc {Chen}, S. \& {Doolen}, G.~D.} 1998 Lattice {B}oltzmann method for fluid
  flows. {\em Annu. Rev. Fluid Mech.\/} {\bf 30}, 329--364.

\bibitem[Chikatamarla \& Karlin(2006)]{Chikatamarla2006}
{\sc Chikatamarla, S.~S. \& Karlin, I.~V.} 2006 Entropy and Galilean
  invariance of lattice {B}oltzmann theories. {\em Phys. Rev. Lett.\/} {\bf
  97}~(19), 190601.

\bibitem[He {\em et~al.\/}(1998)He, Chen \& Doolen]{He1998282}
{\sc He, X.~Y., Chen, S.~Y. \& Doolen, G.~D.} 1998 A novel thermal model for
  the lattice {B}oltzmann method in incompressible limit. {\em J. Comput.
  Phys.\/} {\bf 146}~(1), 282 -- 300.

\bibitem[He \& Luo(1997{\natexlab{{\em a\/}}})]{PhysRevE.55.R6333}
{\sc He, X.~Y. \& Luo, L.~S.} 1997{\natexlab{{\em a\/}}} A priori derivation
  of the lattice {B}oltzmann equation. {\em Phys. Rev. E\/} {\bf 55}~(6),
  R6333--R6336.

\bibitem[He \& Luo(1997{\natexlab{{\em b\/}}})]{PhysRevE.56.6811}
{\sc He, X.~Y. \& Luo, L.~S.} 1997{\natexlab{{\em b\/}}} Theory of the
  lattice {B}oltzmann method: From the {B}oltzmann equation to the lattice
  {B}oltzmann equation. {\em Phys. Rev. E\/} {\bf 56}~(6), 6811--6817.

\bibitem[He {\em et~al.\/}(1996)He, Luo \& Dembo]{He1996}
{\sc He, X.~Y., Luo, L.~S. \& Dembo, M.} 1996 Some progress in lattice
  {B}oltzmann method. part i. nonuniform mesh grids. {\em J. Comput. Phys.\/}
  {\bf 129}~(2), 357--363.

\bibitem[Ho \& Tai(1998)]{annurev.fluid.30.1.579}
{\sc Ho, C.~M. \& Tai, Y.~C} 1998 Micro-electro-mechanical-systems
  ({MEMS}) and fluid flows. {\em Annu. Rev. Fluid Mech.\/} {\bf 30}~(1),
  579--612.

\bibitem[Kim {\em et~al.\/}(2008)Kim, Pitsch \& Boyd]{Kim20088655}
{\sc Kim, S.~H., Pitsch, H. \& Boyd, I.~D.} 2008 Accuracy of
  higher-order lattice {B}oltzmann methods for microscale flows with finite
  {K}nudsen numbers. {\em J. Comput. Phys.\/} {\bf 227}~(19), 8655 -- 8671.

\bibitem[Lim {\em et~al.\/}(2002)Lim, Shu, Niu \& Chew]{lim:2299}
{\sc Lim, C.~Y., Shu, C., Niu, X.~D. \& Chew, Y.~T.} 2002 Application of
  lattice {B}oltzmann method to simulate microchannel flows. {\em Phys.
  Fluids\/} {\bf 14}~(7), 2299--2308.

\bibitem[Lockerby \& Reese(2008)]{Lockerby2008}
{\sc Lockerby, {D.~A.} \& Reese, {J.~M.}} 2008 On the modelling of
  isothermal gas flows at the microscale. {\em J. Fluid Mech.\/} {\bf 604},
  235--261.

\bibitem[Luo(2000)]{Luo200063}
{\sc Luo, L.~S.} 2000 Some recent results on discrete velocity models and
  ramifications for lattice {B}oltzmann equation. {\em Comput. Phys. Commun.\/}
  {\bf 129}~(1-3), 63 -- 74.

\bibitem[Mieussens(2000{\natexlab{{\em a\/}}})]{MIEUSSENS20001}
{\sc Mieussens, L.} 2000{\natexlab{{\em a\/}}} Discrete velocity model and
  implicit scheme for the {BGK} equation of rarefied gas dynamics. {\em Math.
  Models Methods Appl.\/} {\bf 10}, 1121--1149.

\bibitem[Mieussens(2000{\natexlab{{\em b\/}}})]{Mieussens2000}
{\sc Mieussens, L.} 2000{\natexlab{{\em b\/}}} Discrete-velocity models and
  numerical schemes for the {B}oltzmann-{BGK} equation in plane and
  axisymmetric geometries. {\em J. Comput. Phys.\/} {\bf 162}~(2), 429 -- 466.

\bibitem[Mieussens(2001)]{Mieussens200183}
{\sc Mieussens, L.} 2001 Convergence of a discrete-velocity model for the
  {B}oltzmann-{BGK} equation. {\em Comput. Math. Appl\/} {\bf 41}~(1-2), 83 --
  96.

\bibitem[Naris \& Valougeorgis(2005)]{naris:097106}
{\sc Naris, S. \& Valougeorgis, D.} 2005 The driven cavity flow
  over the whole range of the {K}nudsen number. {\em Phys. Fluids\/} {\bf
  17}~(9), 097106.

\bibitem[{Naris} {\em et~al.\/}(2005){Naris}, {Valougeorgis}, {Kalempa} \&
  {Sharipov}]{2005PhFl...17j0607N}
{\sc {Naris}, S., {Valougeorgis}, D., {Kalempa}, D. \& {Sharipov}, F.} 2005
  {Flow of gaseous mixtures through rectangular microchannels driven by
  pressure, temperature, and concentration gradients}. {\em Phys. Fluids\/}
  {\bf 17}~(10), 100607--12.

\bibitem[Nie {\em et~al.\/}(2002)Nie, Doolen \& Chen]{Nie2002}
{\sc Nie, X.~B., Doolen, G.~D. \& Chen, S.~Y.} 2002 Lattice-{B}oltzmann
  simulations of fluid flows in mems. {\em J. Stat. Phys.\/} {\bf 107}~(1),
  279--289.

\bibitem[Qian {\em et~al.\/}(1992)Qian, D'Humi\`{e}res \& Lallemand]{qian1992}
{\sc Qian, Y.~H., D'Humi\`{e}res, D. \& Lallemand, P.} 1992 Lattice {BGK}
  models for {N}avier-{S}tokes equation. {\em Europhys. Lett.\/} {\bf 17}~(6),
  479--484.

\bibitem[Reider \& Sterling(1995)]{Reider1995459}
{\sc Reider, M.~B. \& Sterling, J.~D.} 1995 Accuracy of discrete-velocity
  {BGK} models for the simulation of the incompressible {N}avier-{S}tokes
  equations. {\em Comput. Fluids\/} {\bf 24}~(4), 459 -- 467.

\bibitem[Sbragaglia \& Succi(2005)]{sbragaglia:093602}
{\sc Sbragaglia, M. \& Succi, S.} 2005 Analytical calculation of slip flow in
  lattice {B}oltzmann models with kinetic boundary conditions. {\em Phys.
  Fluids\/} {\bf 17}~(9), 093602.

\bibitem[Sbragaglia \& Succi(2006)]{Sbragaglia2006}
{\sc Sbragaglia, M. \& Succi, S.} 2006 A note on the lattice {B}oltzmann method
  beyond the {C}hapman-{E}nskog limits. {\em Europhys. Lett.\/} {\bf 73}~(3),
  370--376.

\bibitem[Shan \& He(1998)]{PhysRevLett.80.65}
{\sc Shan, X.~W. \& He, X.~Y} 1998 Discretization of the velocity space in
  the solution of the {B}oltzmann equation. {\em Phys. Rev. Lett.\/} {\bf
  80}~(1), 65--68.

\bibitem[{Shan} {\em et~al.\/}(2006){Shan}, {Yuan} \&
  {Chen}]{2006JFM...550..413S}
{\sc {Shan}, X.~W., {Yuan}, X.~F. \& {Chen}, H.~D.} 2006 {Kinetic theory
  representation of hydrodynamics: a way beyond the Navier Stokes equation}.
  {\em J. Fluid Mech.\/} {\bf 550}, 413--441.

\bibitem[Sharipov \& Bertoldo(2009)]{Sharipov2009}
{\sc Sharipov, F. \& Bertoldo, G.} 2009 Numerical solution of the
  linearized {B}oltzmann equation for an arbitrary intermolecular potential.
  {\em J. Comput. Phys.\/} {\bf 228}~(9), 3345 -- 3357.

\bibitem[Sharipov \& Kalempa(2008)]{Sharipov2008}
{\sc Sharipov, F. \& Kalempa, D.} 2008 Oscillatory couette flow at
  arbitrary oscillation frequency over the whole range of the knudsen number.
  {\em Microfluid. Nanofluid.\/} {\bf 4}~(5), 363--374.

\bibitem[Sterling \& Chen(1996)]{Sterling1996}
{\sc Sterling, J.~D. \& Chen, S.~Y.} 1996 Stability analysis of lattice
  {B}oltzmann methods. {\em J. Comput. Phys.\/} {\bf 123}~(1), 196--206.

\bibitem[Struchtrup \& Torrilhon(2003)]{struchtrup:2668}
{\sc Struchtrup, H. \& Torrilhon, M.} 2003 Regularization of {G}rad's
  13 moment equations: Derivation and linear analysis. {\em Phys. Fluids\/}
  {\bf 15}~(9), 2668--2680.

\bibitem[Tang {\em et~al.\/}(2008)Tang, Zhang \& Emerson]{tang:046701}
{\sc Tang, G.~H., Zhang, Y.~H. \& Emerson, D.~R.} 2008 Lattice
  {B}oltzmann models for nonequilibrium gas flows. {\em Phys. Rev. E\/} {\bf
  77}~(4), 046701.

\bibitem[Toschi \& Succi(2005)]{Toschi2005}
{\sc Toschi, F. \& Succi, S.} 2005 Lattice {B}oltzmann method at finite
  {K}nudsen numbers. {\em Europhys. Lett.\/} {\bf 69}~(4), 549--555.

\bibitem[Valougeorgis(1988)]{valougeorgis:521}
{\sc Valougeorgis, D.} 1988 Couette flow of a binary gas mixture. {\em
  Phys. Fluids\/} {\bf 31}~(3), 521--524.

\bibitem[Yang \& Huang(1995)]{Yang1995323}
{\sc Yang, J.~Y. \& Huang, J.~C.} 1995 Rarefied flow computations using
  nonlinear model {B}oltzmann equations. {\em J. Comput. Phys.\/} {\bf
  120}~(2), 323 -- 339.

\bibitem[Yudistiawan {\em et~al.\/}(2008)Yudistiawan, Ansumali \&
  Karlin]{yudistiawan:016705}
{\sc Yudistiawan, W.~P., Ansumali, S. \& Karlin, I.~V.} 2008
  Hydrodynamics beyond {N}avier-{S}tokes: {T}he slip flow model. {\em Phys.
  Rev. E\/} {\bf 78}~(1), 016705.

\bibitem[Zhang {\em et~al.\/}(2005)Zhang, Qin \& Emerson]{zhang:047702}
{\sc Zhang, Y.~H., Qin, R.~S. \& Emerson, D.~R.} 2005 Lattice
  {B}oltzmann simulation of rarefied gas flows in microchannels. {\em Phys.
  Rev. E\/} {\bf 71}~(4), 047702.

\bibitem[Zhang {\em et~al.\/}(2006)Zhang, Gu, Barber \& Emerson]{zhang:046704}
{\sc Zhang, Y.~H., Gu, X.~J., Barber, R.~W. \& Emerson, D.~R.}
  2006 Capturing {K}nudsen layer phenomena using a lattice {B}oltzmann model.
  {\em Phys. Rev. E\/} {\bf 74}~(4), 046704.

\bibitem[{Zhong} {\em et~al.\/}(1993){Zhong}, {Chapman} \&
  {MacCormack}]{1993aiaaj..31.1036z}
{\sc {Zhong}, X.~L., {Chapman}, D.~R. \& {MacCormack}, R.~W.} 1993 Stabilization
  of the {B}urnett equations and application to hypersonicflows. {\em AIAA
  Journal\/} {\bf 31}, 1036--1043.

\end{thebibliography}
\end{document}